\documentclass[journal]{IEEEtran}

\usepackage{amsthm}
\usepackage{graphicx}
\usepackage{booktabs}
\usepackage{cite}
\usepackage{bbding}
\usepackage{amsmath}
\usepackage{amsfonts}
\usepackage{cases}
\usepackage{algorithmic}
\usepackage{algorithm}
\usepackage{array}
\usepackage{multirow}
\usepackage{threeparttable}
\usepackage{bm}
\usepackage[colorlinks,linkcolor=black, citecolor=blue]{hyperref}
\usepackage[version=4]{mhchem}
\usepackage{amsmath,nccmath}

\begin{document}

\title{ A Guaranteed Convergence Analysis for the Projected Fast Iterative Soft-Thresholding Algorithm in Parallel MRI}

\author{Xinlin~Zhang,
        Hengfa~Lu,
        Di~Guo,
        Lijun~Bao,
        Feng~Huang,
        Qin~Xu,
        Xiaobo~Qu*
\thanks{This work was supported in part by National Key R\&D Program of China (2017YFC0108703), National Natural Science Foundation of China (61971361, 61871341, 61811530021, U1632274 and 61672335), Natural Science Foundation of Fujian Province of China (2018J06018), Fundamental Research Funds for the Central Universities (20720180056), Science and Technology Program of Xiamen (3502Z20183053), and Xiamen University Nanqiang Outstanding Talents Program. \emph{*Corresponding author: Xiaobo Qu}.}%
\thanks{Xinlin Zhang, Hengfa Lu, Lijun Bao and Xiaobo Qu are with Department of Electronic Science, Fujian Provincial Key Laboratory of Plasma and Magnetic Resonance, School of Electronic Science and Engineering, National Model Microelectronics College, Xiamen University, Xiamen 361005, China (e-mail: quxiaobo@xmu.edu.cn).}
\thanks{Di Guo is with School of Computer and Information Engineering, Xiamen University of Technology, Xiamen 361024, China.}%
\thanks{Feng Huang and Qin Xu are with Neusoft Medical System, Shanghai 200241, China.}}

% \markboth{IEEE TRANSACTIONS ON COMPUTATIONAL IMAGING}%
% {Shell \MakeLowercase{\textit{et al.}}: Bare Demo of IEEEtran.cls for IEEE Journals}

\maketitle

%----------------------------------------------------------------------
%----------------------------- Abstract -------------------------------
%----------------------------------------------------------------------
\begin{abstract}
The boom of non-uniform sampling and compressed sensing techniques dramatically alleviates the lengthy data acquisition problem of magnetic resonance imaging. Sparse reconstruction, thanks to its fast computation and promising performance, has attracted researchers to put numerous efforts on it and has been adopted in commercial scanners. To perform sparse reconstruction, choosing a proper algorithm is essential in providing satisfying results and saving time in tuning parameters. The pFISTA, a simple and efficient algorithm for sparse reconstruction, has been successfully extended to parallel imaging. However, its convergence criterion is still an open question. And the existing convergence criterion of single-coil pFISTA cannot be applied to the parallel imaging pFISTA, which, therefore, imposes confusions and difficulties on users about determining the only parameter - step size.
In this work, we provide the guaranteed convergence analysis of the parallel imaging version pFISTA to solve the two well-known parallel imaging reconstruction models, SENSE and SPIRiT. Along with the convergence analysis, we provide recommended step size values for SENSE and SPIRiT reconstructions to obtain fast and promising reconstructions.
Experiments on \textit{in vivo} brain images demonstrate the validity of the convergence criterion. Besides, experimental results show that compared to using backtracking and power iteration to determine the step size, our recommended step size achieves more than five times acceleration in reconstruction time in most tested cases.
\end{abstract}


\IEEEpeerreviewmaketitle
%----------------------------------------------------------------------
%--------------------------- Introduction -----------------------------
%----------------------------------------------------------------------
\section{Introduction}\label{Section:introduction}
\IEEEPARstart{M}{agnetic} resonance imaging (MRI) is a non-invasive, non-radioactive, and versatile technique serving as a widely adopted and indispensable tool in medical diagnosis. However, the slow imaging speed impedes its development. The advent of sparse sampling and compressed sensing (CS) theory \cite{2006_TIT_CS, 2006_TIT_Tao, 2007_MRM_Sparse_MRI} meets the eager demand of fast scan through sampling only a small amount of data points and recovering the missing data using well-developed reconstruction methods.

Sparsity \cite{2007_MRM_Sparse_MRI,2008_Qu_contourlet,2010_IPSE_Contourlet_MRI,2011_TMI_Redundant_wavelet_recon, 2011_SIDWT_Ying, 2011_TMI_Redundant_Sai,2011_MIA_Orthogonal_1,2012_MRI_PBDW,2014_MIA_PANO,2008_Orthogonal_2}, low rank \cite{2007_PSF_liang,2010_PJO_Low_rank,2015_MRM_Low_rank,2016_liang_LR,2020_xinlin}, and sparsity plus low rank \cite{2011_TMI_k-t_SLR,2015_MRM_L+S,2018_MRM_L+S,2018_TCI_L+S_FISTA} are common-adopted priors used in MRI image reconstruction. In this work, we will focus on sparse reconstructions. Particularly, most emphasis will be put into algorithms to solve sparse reconstruction models. The sparse representation adopted to empower the image to be sparse plays a crucial role in designing a reconstruction approach. To the best of our knowledge, sparse representation approaches could be categorized into two main genres: orthogonal \cite{2007_MRM_Sparse_MRI, 2008_Orthogonal_2, 2011_MIA_Orthogonal_1} and redundant representation systems \cite{2008_Qu_contourlet,2010_IPSE_Contourlet_MRI,2011_TMI_Redundant_wavelet_recon, 2011_SIDWT_Ying, 2011_TMI_Redundant_Sai,  2012_MRI_PBDW, 2016_TBME_FDLCP, 2014_MIA_PANO, 2016_MIA_GBRWT}. Representation systems that can sparsify MRI images include transforms, such as wavelets, dictionaries, adaptive representations, etc.
The redundant representation systems are favored in sparse MRI reconstructions as they enable sparser image representation than the orthogonal representation systems can do, suggesting better noise removal and artifacts suppression in applications.

Redundant representation systems are described by frame, mostly tight frame \cite{2014_tight_frame, 2016_TMI_pFISTA}, leading to two distinct kinds of reconstruction models, the synthetic model \cite{2008_CRM_CS,2006_TIT_Tao,2006_TIT_CS}, and the analysis model \cite{2008_TIT_redundant_CS,2011_ACHA_redundant_CS,2013_ACHA_analysis_model}. The readers are referred to \cite{2016_TMI_pFISTA} for definitions of the tight frame, analysis model, and synthetic model in CS MRI. The analysis model, assuming the coefficients in the transform domain of an image to be sparse, and synthesis model, considering an image as a linear combination of sparse coefficients, have different prior assumptions. Even with the same MRI data, sampling pattern, and sparse transform, the analysis model is observed to yield improved reconstruction results compared to the synthesis model \cite{2015_PlosOne_Balance, 2016_TMI_pFISTA}. Besides, it has been shown that the balanced model lies in between the synthetic model and the analysis model. In the context of MRI reconstruction, Liu et al. empirically explored the performance of the balanced model and observed that the balanced model has a comparable reconstruction performance with the analysis model \cite{2015_PlosOne_Balance}.

Analysis models, though enable better reconstructions with smaller errors, still have a compelling demand for fast algorithms that allow favorable convergence speed and fewer parameters. Many algorithms have been developed to solve the analysis models \cite{2020_jeffery}, such as alternating direction methods of multipliers (ADMM) \cite{2011_ADMM,2010_TMI_AL}, nonlinear conjugate gradient (NLCG) \cite{2007_MRM_Sparse_MRI}, variants of Nesterov's algorithm \cite{1983_NESTA_1,2005_NESTA_2,2013_NESTA_3}, and Douglas Rachford splitting \cite{1956_Douglas-Rachford,2007_JSTSP_Douglas-Rachford_Recovery}. However, they are time- and memory-demanding or vulnerable to parameter selections. The original iterative shrinkage threshold algorithm (ISTA) \cite{2004_ISTA} and its acceleration version - fast ISTA (FISTA) \cite{2009_SIAM_FISTA} are efficient and robust. Nevertheless, they are limited to solve the synthesis model. Numerous efforts have been made by researchers to improve FISTA, providing us many variants of FISTA \cite{2011_TP_variants_FISTA_theory,2018_JOTA_variants_FISTA_theory}. Some variants of FISTA have been proposed to solve analysis models, such as MFISTA-FGP \cite{2009_TIP_FISTA_FGP}, which provides guaranteed non-increasing function values and MFISTA-VA \cite{2018_TCI_MFISTA_VA} which utilizes variables acceleration to achieve faster convergence speed while keeping the monotonality. However, they are either computational demanding or have more than one parameter to tune. In contrast, a variant of FISTA developed by our group named projected iterative soft-threshold algorithm (pISTA) and its acceleration version - pFISTA \cite{2016_TMI_pFISTA}, enables faster reconstructions, requires less memory space and has only one adjustable parameter, the step size $\gamma$. Moreover, its convergence criterion has been provided in the paper \cite{2016_TMI_pFISTA}. Also, Liu et al. \cite{2016_TMI_pFISTA} converted the analysis model into an equivalent synthesis-like one with a constraint on its range, which then solved by orthogonal projection, and theoretically proved that the pFISTA converges to a balanced model.

The pFISTA, however, has limitations, such as it could not deal with the total variation (TV), and it is limited to tackle single-coil image reconstruction problems. To solve the single-coil problem, Ting et al. independently proposed a computationally efficient balanced sparse reconstruction method in the context of parallel MRI under tight frame\cite{2017_MRM_bFISTA,2020_Ahmad}, named bFISTA, and applied bFISTA to two widely adopted parallel imaging models, sensitivity encoding (SENSE) method \cite{1999_MRM_SENSE} and iterative self-consistent parallel imaging reconstruction (SPIRiT) \cite{2010_MRM_SPIRiT}. However, the authors did not provide proof of the convergence of bFISTA; that is to say, in practice, there is no guidance about how to choose $\gamma$. In addition, the convergence criterion proved for the single-coil pFISTA cannot be directly applied to the multi-coil cases. Therefore, the algorithm users would encounter a problem of how to choose a proper $\gamma$ to produce faithful results. We give an example in Fig. \ref{fig_RelatedWork} (Section \ref{Subsection:relatedWork_pFISTA-parallel}) to demonstrate this issue. Besides, backtracking and power iteration, though being useful to calculate the $\gamma$, are time-consuming (Section \ref{Section:experimentalResults}). Due to the importance of parallel imaging, it is necessary to give a clear mathematical proof of its convergence to assist in setting a proper $\gamma$.

In this work, we provide sufficient conditions for the convergence of parallel imaging version pFISTA and explicitly provide convergence criteria of applying pFISTA on solving two exemplars of parallel imaging reconstruction methods - SENSE and SPIRiT. With the convergence analysis, recommend $\gamma$ for SENSE and SPIRiT reconstructions using pFISTA to permit the fastest convergence speed and promising results. We first assess the influence of the gap between the recommended $\gamma$ and hand-tuned optimal $\gamma$ on the convergence speed. Then we compare our method with backtracking and power iteration, in which our recommended $\gamma$ is much faster and enables reliable reconstructions. Also, we compare other variants of FISTA, ADMM, NLCG, and pFISTA. Furthermore, we discuss the results of applying pFISTA on parallel reconstruction models under different tight frames.

The rest of the paper is organized as follows. In Section \ref{Section:notations}, we introduce the notations. In Section \ref{Section:relatedWork}, we introduce some related works, firstly the pFISTA, and then SENSE and SPIRiT. In section \ref{Section:convergence}, we prove that the parallel imaging version pFISTA converges under a proper selection of the step size. Furthermore, we offer the convergence criteria of pFISTA when applied to tackle SENSE and SPIRiT models. In Section \ref{Section:experimentalResults}, we demonstrate the usefulness of the criteria we provided with multiple parallel imaging brain images. Finally, conclusions will be drawn in Section \ref{Section:conclusion}.%

%----------------------------------------------------------------------
%------------------------------ Notations -----------------------------
%----------------------------------------------------------------------
\section{Notations}\label{Section:notations}
We first introduce notations used throughout this paper. We denote vectors by bold lowercase letters and matrices by bold uppercase letters. The transpose and conjugate transpose of a matrix are denoted by  $\mathbf{X}^T$ and  $\mathbf{X}^H$. For any vector $\mathbf{x}$, $\left\| \mathbf{x} \right\|_1$ and $\left\| \mathbf{x} \right\|_2$ denote the $\ell_1$ and $\ell_2$ norm for vectors, respectively. For a matrix $\mathbf{X}$, $\left\| \mathbf{X} \right\|_2$ denotes the $\ell_2$ norm for matrix, which is the largest singular value of matrix $\mathbf{X}$ and also the square root of the largest eigenvalue of the matrix $\mathbf{X}^H \mathbf{X}$. 

Operators are denoted by calligraphic letters. Let ${\cal D}_M$ denotes block diagonalization operator which places any $M$ matrices of the same size, $\mathbf{X}_1, \cdots, \mathbf{X}_M$, along with the diagonal entries of a matrix with zeros:
\begin{equation}\label{(D)}
{\cal D}_M \left( \mathbf{X}_{1}, \cdots, \mathbf{X}_{M} \right) = \left[ \begin{matrix}
   \mathbf{X}_1 & {} & \mathbf{0}  \\
   {} & \ddots  & {}  \\
   \mathbf{0} & {} & \mathbf{X}_M  \\
\end{matrix} \right].
\end{equation}

%----------------------------------------------------------------------
%---------------------------- Related Work ----------------------------
%----------------------------------------------------------------------
\section{Related Work} \label{Section:relatedWork}
\subsection{pFISTA for Single-Coil MRI Reconstruction}
An analysis model for single-coil sparse MRI reconstruction could be formulated as
\begin{equation} \label{(single-coil model)}
	\mathop {\min }\limits_{{{\mathbf{x}}_s}} \lambda {\left\| {{\mathbf{\Psi }}{{\mathbf{x}}_s}} \right\|_1} + \frac{1}{2}\left\| {{{\mathbf{y}}_s} - {\mathbf{UF}}{{\mathbf{x}}_s}} \right\|_2^2,
\end{equation}
where ${{\mathbf{x}}_{s}}\in {{\mathbb{C}}^{N}}$ denotes the single-coil MR image data rearranged into a column vector, ${{\mathbf{y}}_{s}}\in {{\mathbb{C}}^{M}}$ the single-coil undersampled k-space data, $\mathbf{U}\in {{\mathbb{R}}^{M\times N}}\left( M \ll N \right)$ the undersampling matrix, and $\mathbf{F}\in {{\mathbb{C}}^{N\times N}}$ the discrete Fourier transform. $\mathbf{\Psi }$ is a tight frame, and the constant $\lambda $ is the regularization parameter to balance the sparsity and data consistency.

To solve the problem \eqref{(single-coil model)}, pFISTA rewrites the formula mentioned above as a synthetic model as
\begin{equation}\label{(synthetic model)}
  \mathop {\min }\limits_{{\bm{\alpha }} \in {\rm{Range}}\left( {\mathbf{\Psi }} \right)} \lambda {\left\| {\bm{\alpha }} \right\|_1} + \frac{1}{2}\left\| {{{\mathbf{y}}_s} - {\mathbf{UF\Psi} ^{*} \bm{\alpha}}} \right\|_2^2,
\end{equation}
where ${{\mathbf{\Psi }}^{*}}$ denotes the adjoint of $\mathbf{\Psi }$, and specifically satisfies ${{\mathbf{\Psi }}^{*}} \mathbf{\Psi } =\mathbf{I}$. $\bm{\alpha }$ contains the coefficients of an image under the representation of a tight frame $\mathbf{\Psi}^{*}$.

According to \cite{2016_TMI_pFISTA}, the main iterations of pFISTA to solve the problem in Eq. \eqref{(synthetic model)} are
\begin{equation}\label{single-coil solution}
% \begin{medsize}
  \begin{aligned}
    {\mathbf{x}}_s^{\left( {k + 1} \right)} & \! = \! {{\mathbf{\Psi }}^*}{T_{\gamma \lambda }} \! \left( \! {{\mathbf{\Psi }} \! \left( {\hat{\mathbf{x}}_s^{\left( k \right)} \! + \! \gamma {{\mathbf{F}}^H}{{\mathbf{U}}^T} \! \left( \! {{{\mathbf{y}}_s} \! - \! {\mathbf{UF}\hat{\mathbf{x}}}_s^{\left( k \! \right)}} \right)} \! \right)} \! \right),\\
    {t^{\left( {k + 1} \right)}} & \! = \! \frac{{1 + \sqrt {1 + 4{{\left( {{t^{\left( k \right)}}} \right)}^2}} }}{2},\\
    {\hat{\mathbf{x}}}_s^{\left( {k + 1} \right)} & \! = \! {\mathbf{x}}_s^{\left( {k + 1} \right)} + \frac{{{t^{\left( {k} \right)}} - 1}}{t^{\left( {k + 1} \right)}}\left( {{\mathbf{x}}_s^{\left( {k + 1} \right)} - {\mathbf{x}}_s^{\left( k \right)}} \right),
\end{aligned}
% \end{medsize}
\end{equation}
where ${{T}_{\gamma \lambda }}\left( \cdot  \right)$ is a point-wise soft-thresholding function defined as ${{T}_{\gamma \lambda }}\left( {\alpha}  \right)=\max \left\{ \left| {\alpha}  \right|-\gamma \lambda ,0 \right\}\cdot {\alpha} / {\left| {\alpha}  \right|}$.

According to Theorem 2 in the pFISTA paper \cite{2016_TMI_pFISTA}, when the step size $0<\gamma \le 1$, the algorithm will converge. Besides, the larger $\gamma $ is, the faster pFISTA converges. Therefore, $\gamma =1$ is recommended in pFISTA to produce promising reconstruction with the fastest convergence speed.

% It is worthy to notice that starting from an analysis model, pFISTA first converts the analysis model into a synthetic-like model with a constraint on the range. Notably, the two models are equivalent \cite{2016_TMI_pFISTA}. Then the synthetic-like model with ${\bm{\alpha }} \in {\rm{Range}}\left( {\mathbf{\Psi }} \right)$ constraint was solved with orthogonal projection. The pFISTA coincide to converges to a balanced model with an $\ell_2$-type penalty \cite{2016_TMI_pFISTA}.

\subsection{pFISTA for Multi-Coil MRI Reconstruction}\label{Subsection:relatedWork_pFISTA-parallel}
According to \cite{2017_MRM_bFISTA}, we can formulate analysis models for the parallel MRI reconstruction problem into a unified form as
\begin{equation} \label{(pFISTA-parallel model)}
        \small{\left ( \textbf{pFISTA-parallel} \right )} \quad \mathop {\min }\limits_{\mathbf{d}} \lambda {\left\| {{\mathbf{\Psi d}}} \right\|_1} + \left\| {{\mathbf{y}} - {\mathbf{Ad}}} \right\|_2^2,
\end{equation}
where $\mathbf{d}$ represents the desired image to be recovered, $\mathbf{y}=\left[ {{\mathbf{y}}_{1}};{{\mathbf{y}}_{2}};\cdots ;{{\mathbf{y}}_{J}} \right] \in{\mathbb{C}^{MJ}}$ the undersampled multi-coil k-space data rearranged into a column vector, and ${{\mathbf{y}}_{j}}\in{\mathbb{C}^{M}} \;\left( j=1,2,\cdots ,J \right)$ is the undersampled k-space data vector of $j^{th}$ coil, and $\mathbf{A}$ as a system matrix in parallel MRI, including coils modulation, Fourier transform, and undersampling.

For parallel MRI reconstruction methods based on different signal properties, the explicit expressions of Eq. \eqref{(pFISTA-parallel model)} would vary. Two reconstruction algorithms based on SENSE and SPIRiT are discussed in \cite{2017_MRM_bFISTA}. However, the convergence of these two algorithms has not been proven. Thus, in this work, we first prove the convergence of pFISTA of solving the general parallel MRI reconstruction model and then offer two concrete examples of multi-coils MRI analysis model, SENSE and SPIRiT, with convergence analysis. We first introduce how to solve SENSE and SPIRiT using pFISTA.

\begin{figure}[htb]
\setlength{\abovecaptionskip}{0.cm}
\setlength{\belowcaptionskip}{-0.cm}
\centering
\includegraphics[width=3in]{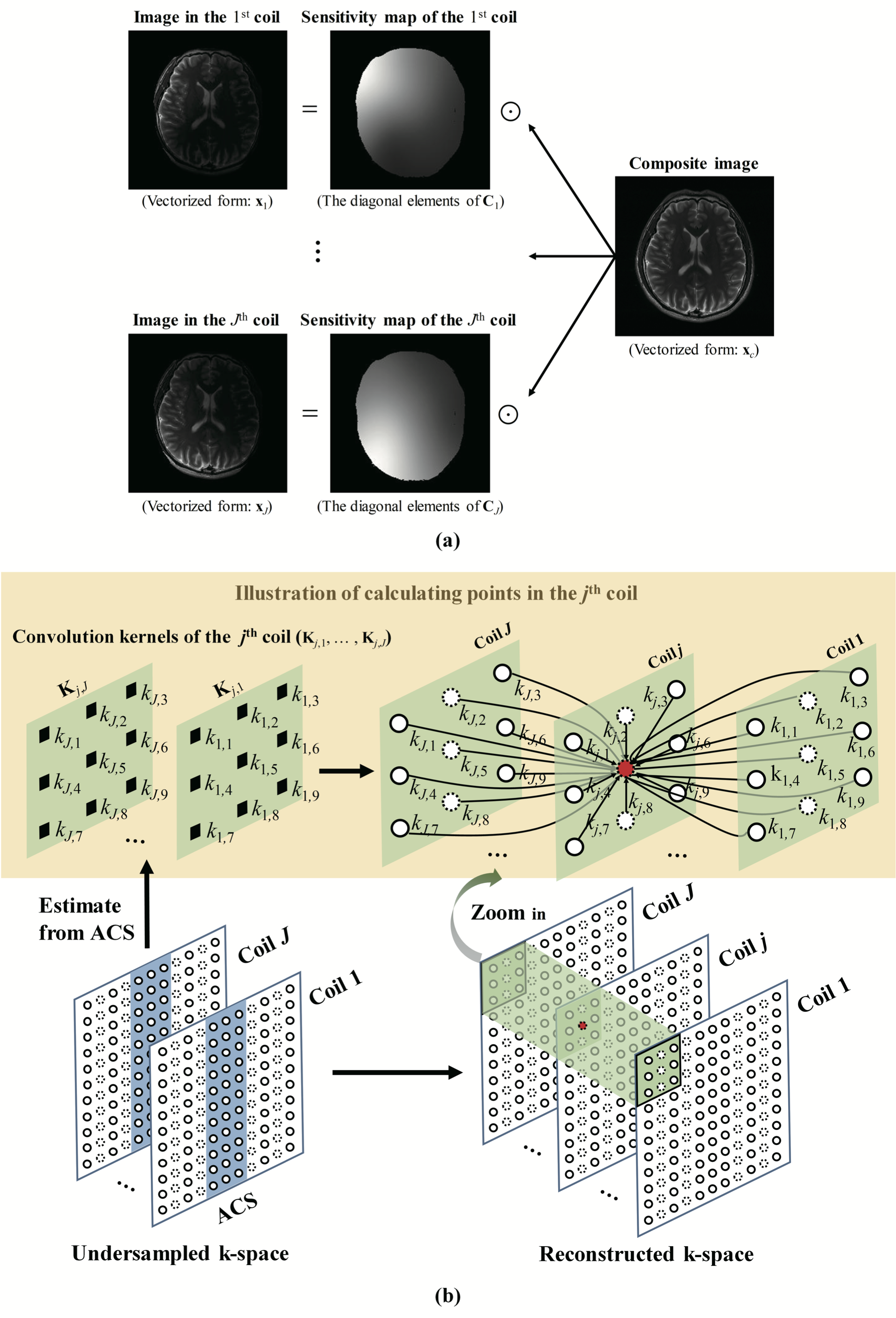}
\caption{Parallel imaging reconstruction methods. (a) SENSE; (b) SPIRiT. Here $\odot$ denotes Hadamard product.}
\label{fig_SENSE_SPIRiT}
\end{figure}

\subsubsection{pFISTA-SENSE}
As shown in Fig. \ref{fig_SENSE_SPIRiT} (a), in SENSE \cite{1999_MRM_SENSE}, the image ${{\mathbf{x}}_{j}}\in {{\mathbb{C}}^{N}}$ of the $j^{th}$ coil is represented as:
\begin{equation}\label{(SENSE)}
  {{\mathbf{x}}_j} = {{\mathbf{C}}_j}{{\mathbf{x}}_{c}},\;j = 1,2,...,J,
\end{equation}
where ${{\mathbf{x}}_{j}}$ and ${{\mathbf{x}}_{c}}\in {{\mathbb{C}}^{N}}$ denote the $j^{th}$ coil image and the composite MRI image rearranged into a column vector, ${{\mathbf{C}}_{j}}\in {{\mathbb{C}}^{N\times N}},\left( j=1,2,...,J \right)$ is a diagonal matrix which contains the sensitivity map of the $j^{th}$ coil.

The reconstruction problem based on SENSE can be formulated as:
\begin{equation} \label{(pFSITA-SENSE model)}
        \small{{\left ( \textbf{pFISTA-SENSE} \right )}} \quad \mathop {\min }\limits_{{{\mathbf{x}}_{c}}} \lambda {\left\| {{\mathbf{\Psi }}{{\mathbf{x}}_{c}}} \right\|_1} + \frac{1}{2}\left\| {{\mathbf{y}} - {{\tilde{\mathbf{U}}}{\tilde{\mathbf{F}}}\mathbf{C}{{\mathbf{x}}_{c}}}} \right\|_2^2,
\end{equation}
where $\tilde{\mathbf{U}}={\cal D}_J \left( \mathbf{U}, \cdots, \mathbf{U} \right)$, $\tilde{\mathbf{F}}={\cal D}_J \left( \mathbf{F}, \cdots, \mathbf{F} \right)$, $\mathbf{C}=\left[ {{\mathbf{C}}_{1}};{{\mathbf{C}}_{2}};\cdots ;{{\mathbf{C}}_{J}} \right]\in {{\mathbb{C}}^{NJ\times N}}$,  Here, the system matrix $\mathbf{A}$ in Eq. \eqref{(pFISTA-parallel model)} has its explicit expression as $\mathbf{A}={\tilde{\mathbf{U}}}{\tilde{\mathbf{F}}}\mathbf{C}$.

Using pFISTA, we can get the solution of Eq. \eqref{(pFSITA-SENSE model)} by iteratively solving the following problems:
\begin{equation}\label{(pFISTA-SENSE solution)}
\begin{medsize}
  \begin{aligned}
    {\mathbf{x}}_{c}^{\left( {k + 1} \right)} &= {{\mathbf{\Psi }}^*}{T_{\gamma \lambda }}\left( {{\mathbf{\Psi }}\left( {{\hat{\mathbf x}}_{c}^{\left( k \right)} \! + \! \gamma {{\mathbf{C}}^H}{{{\tilde{\mathbf F}}}^H}{{{\tilde{\mathbf U}}}^T}\left( {{\mathbf{y}} \! - \! \tilde{\mathbf U}\tilde{\mathbf{F}}\mathbf{C}\hat{\mathbf{x}}_{c}^{\left( k \right)}} \right)} \right)} \right),\\
    {t^{\left( {k + 1} \right)}} &= \frac{{1 + \sqrt {1 + 4{{\left( {{t^{\left( k \right)}}} \right)}^2}} }}{2},\\
    {\hat{\mathbf x}}_{c}^{\left( {k + 1} \right)} &= {\mathbf{x}}_{c}^{\left( {k + 1} \right)} + \frac{{{t^{\left( {k } \right) }} - 1}}{t^{\left( {k + 1} \right) }}\left( {{\mathbf{x}}_{c}^{\left( {k + 1} \right)} - {\mathbf{x}}_{c}^{\left( k \right)}} \right).
    \end{aligned}
    \end{medsize}
\end{equation}

For simplicity, we call the pFISTA adopted to solve SENSE as pFISTA-SENSE.

\subsubsection{pFISTA-SPIRiT}
The SPIRiT \cite{2010_MRM_SPIRiT} primarily bases on the assumption that each k-space data point of a given coil is the convolution of the multi-coil data of its neighboring k-space points, and the convolution kernels are estimated from auto-calibration signal (ACS) (Fig. \ref{fig_SENSE_SPIRiT} (b)). Let $\mathbf{x}=\left[ {{\mathbf{x}}_{1}};{{\mathbf{x}}_{2}};\cdots ;{{\mathbf{x}}_{J}} \right]\in {{\mathbb{C}}^{NJ}}$ denote the multi-coil image data rearranged into a column vector, where ${{\mathbf{x}}_{j}}\in {{\mathbb{C}}^{N}},\left( j=1,2,\cdots J \right)$ is the ${{j}^{th}}$ coil image vector, then the calibration consistency in image domain SPIRiT can be formulated as:

\begin{equation}\label{(SPIRiT-W)}
{{\mathbf{x}}_{j}}=\left[ {{\mathbf{W}}_{j,1}},{{\mathbf{W}}_{j,2}},\cdots {{\mathbf{W}}_{j,J}} \right]\mathbf{x},
\end{equation}
where ${{\mathbf{W}}_{j,i}}\in {{\mathbb{C}}^{N\times N}}\ \left( i=1,2,\cdots ,J \right)$ is a diagonal matrix with the diagonal elements being the inverse Fourier transform of the convolution kernel $\mathbf{K}_{j,i}$ in the Fig. \ref{fig_SENSE_SPIRiT} (b). Then, the $\ell_{1}$-SPIRiT reconstruction can be formulated as:
\begin{equation}\label{(pFISTA-SPIRiT model)}
% \begin{medsize}
        \begin{array}{l}
        \small{\left ( \text{\textbf{pFISTA-SPIRiT}} \right )}  \\
        \mathop {\min }\limits_{\mathbf{x}} \lambda {\left\| {{\mathbf{\Psi x}}} \right\|_1} \! + \! \frac{1}{2} \left\| {{\mathbf{y}} - {\tilde{\mathbf U}\tilde{\mathbf{F}}\mathbf{x}}} \right\|_2^2 \! + \! \frac{\lambda _1}{2} \left\| {\left( {{\mathbf{W}} - {\mathbf{I}}} \right){\mathbf{x}}} \right\|_2^2,
         \end{array}
% \end{medsize}
\end{equation}
where the matrix $\mathbf{W}\in {{\mathbb{C}}^{NJ\times NJ}}$ is
\begin{equation}\label{(W)}
\mathbf{W}=\left[ \begin{matrix}
   {{\mathbf{W}}_{1,1}} & {{\mathbf{W}}_{1,2}} & \cdots  & {{\mathbf{W}}_{1,J}}  \\
   {{\mathbf{W}}_{2,1}} & {{\mathbf{W}}_{2,2}} & \cdots  & {{\mathbf{W}}_{2,J}}  \\
   \vdots  & \vdots  & \ddots  & \vdots   \\
   {{\mathbf{W}}_{J,1}} & {{\mathbf{W}}_{J,2}} & \cdots  & {{\mathbf{W}}_{J,J}}  \\
\end{matrix} \right].
\end{equation}
Notice that strictly speaking, the $\mathbf{\Psi}$ in Eq. \eqref{(pFISTA-SPIRiT model)} should be wrote in the form of $\tilde{\mathbf{\Psi }}={\cal D}_J \left( \mathbf{\Psi}, \cdots, \mathbf{\Psi} \right)$ indicating that the $\mathbf{\Psi }$ is applied to each coil image. Here $\tilde{\mathbf{\Psi }}$ is still a tight frame which satisfies ${\tilde{{\mathbf{\Psi }}}^{*}} \tilde{\mathbf{\Psi }}=\mathbf{I}$, we use only $\mathbf{\Psi }$ in the rest of the paper for simplicity.

We reformulate Eq. \eqref{(pFISTA-SPIRiT model)} to line up with Equation \eqref{(pFISTA-parallel model)} as:
\begin{equation}\label{(pFISTA-SPIRiT model-1)}
% \begin{medsize}
        \begin{array}{l}
        \small{\left ( \text{\textbf{pFISTA-SPIRiT}} \right )}  \\
        \mathop {\min }\limits_{\mathbf{x}} \lambda {\left\| {{\mathbf{\Psi x}}} \right\|_1} \! + \! \frac{1}{2} \! \left\| {\left[ \! \begin{array}{l}
        {\mathbf{I}}\\
        {\mathbf{0}}
        \end{array} \! \right]{\mathbf{y}} \! - \! \left[ \! \! {\begin{array}{*{20}{c}}
        {{\tilde{\mathbf U}}{\tilde{\mathbf F}}}\\
        { - \sqrt {{\lambda _1}} \left( {{\mathbf{W}} - {\mathbf{I}}} \right)}
        \end{array}} \! \! \right] \! {\mathbf{x}}} \right\|_2^2,
         \end{array}
% \end{medsize}
\end{equation}
Here, the system matrix $\mathbf{A}$ has its explicit expression as ${\bf{A}} = \left[ {\begin{array}{*{20}{c}}
{\tilde{\mathbf{U}}}{\tilde{\mathbf{F}}}&- \sqrt {{\lambda _1}} \left( {{\mathbf{W}} - {\mathbf{I}}} \right)
\end{array}} \right]^{T}$.

Using pFISTA, we can get the solution of Eq. \eqref{(pFISTA-SPIRiT model-1)} by iteratively solving the following problems:
\begin{equation}\label{(pFISTA-SPIRiT solution)}
\begin{medsize}
\begin{aligned}
     {\mathbf{x}^{\left( {k+1} \right)}} &= {{\mathbf{\Psi }}^*} {T_{\gamma \lambda }} \left( {\mathbf{\Psi }} \left( {\hat{\mathbf x}}^{\left( k \right)} + \gamma \left( {\tilde{\mathbf{F}}^H}{{\tilde{\mathbf U}}^T}  \left( \mathbf{y} -  {\tilde{\mathbf U}} {\tilde{\mathbf F}} {\hat{\mathbf{x}}^{\left( k \right)}}\right) \right. \right. \right.\\
     & \;\;\; \left. \left. \left.  - {\lambda _1} {\left( \mathbf{W}-\mathbf{I} \right)}^{H}  {\left( \mathbf{W}-\mathbf{I} \right)}  {\hat{\mathbf{x}}} \right) \right) \right),\\
      {t^{\left( {k + 1} \right)}}  &= \frac{{1 + \sqrt {1 + 4{{\left( {{t^{\left( k \right)}}} \right)}^2}} }}{2},\\
      {{{\hat{\mathbf x}}}^{\left( {k + 1} \right)}} &= {{\mathbf{x}}^{\left( {k + 1} \right)}} + \frac{{{t^{\left( k \right)}} - 1}}{{{t^{\left( {k + 1} \right)}}}}\left( {{{\mathbf{x}}^{\left( {k + 1} \right)}} - {{\mathbf{x}}^{\left( k \right)}}} \right).
\end{aligned}
\end{medsize}
\end{equation}

\subsection{Connection Between pFISTA and A Balanced Model}
Starting from an analysis model, pFISTA first converts the analysis model into a synthetic-like model with a constraint $\mathbf{\alpha }\in \text{Range}\left( \mathbf{\Psi } \right)$. Notably, the two models are equivalent \cite{ 2016_TMI_pFISTA}. Then the synthetic-like model with a constraint on the range was solved with orthogonal projection. 

Importantly, pFISTA happens to converge to a balanced model with an $\ell_2$-type penalty $ \frac{1}{2\gamma }\left\| \left( \mathbf{I}-\mathbf{\Psi }{{\mathbf{\Psi }}^{*}} \right)\mathbf{\alpha } \right\|_{2}^{2}$. Despite that FISTA can solve a general balanced model, the importance of pFISTA is not hampered because, first, pFISTA introduces only one parameter $\gamma$, and experimental results showed that pFISTA reconstructions are robust to $\gamma$; second, pFISTA is more memory-efficient as it performs reconstruction in the image domain rather than the coefficients domain where FISTA reconstructs the signal. FISTA has to allocate considerable memory spaces to store the redundant coefficients.

\begin{figure}[!h]
\setlength{\abovecaptionskip}{0.cm}
\setlength{\belowcaptionskip}{-0.cm}
\centering
\includegraphics[width=3.2in]{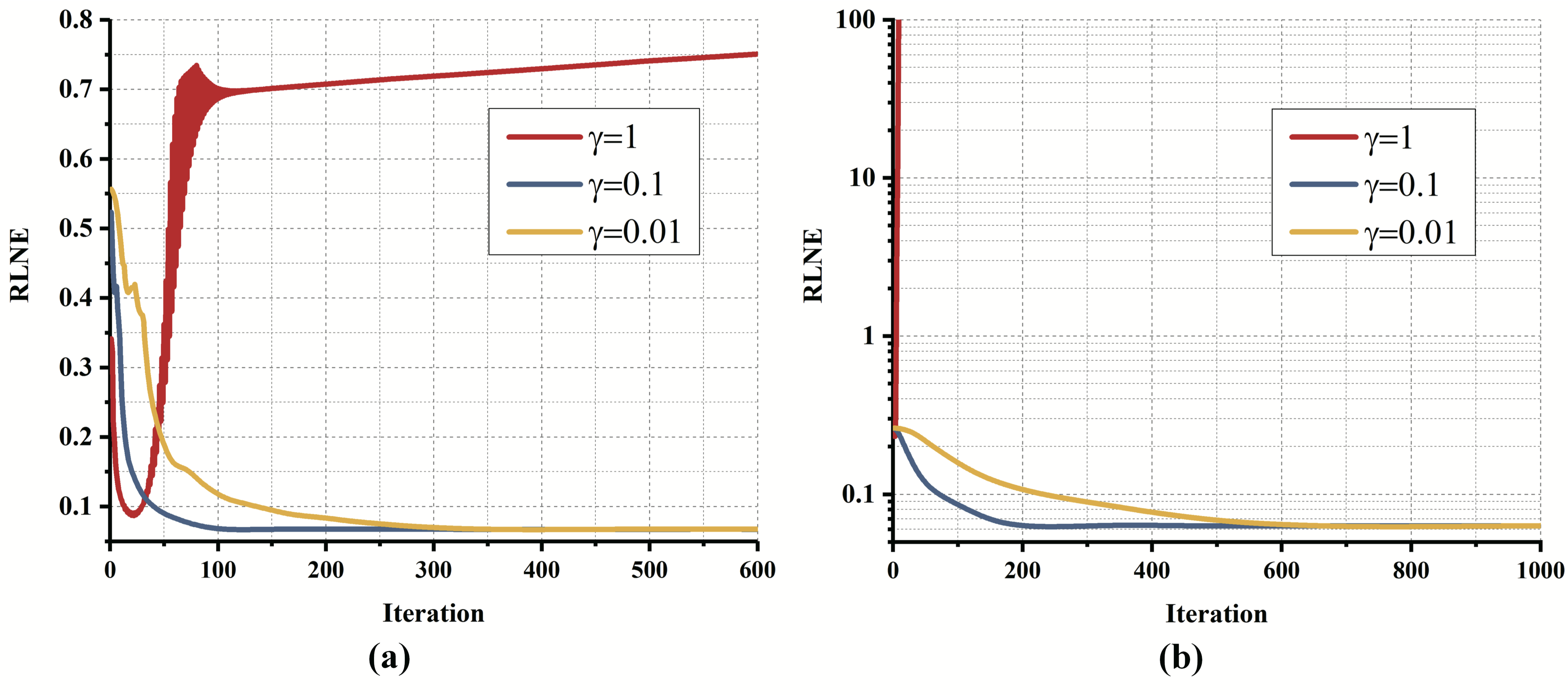}
\caption{Empirical convergence of pFISTA-SENSE (a) and pFISTA-SPIRiT (b) with different $\gamma$. The reconstruction experiments were carried out on a 32-coil brain image with 34\% data acquired using a Cartesian sampling pattern. The sensitivity matrix $\mathbf{C}$ was normalized using its max absolute value. The tested data and the sampling pattern are presented in Fig. \ref{fig_Dataset}.}
\label{fig_RelatedWork}
\end{figure}

It is worthy to point out that convergence analysis plays an important role for users in determining the parameter to produce promising results. Despite pFISTA coincides with FISTA if the weighting parameter of the term $\left\| \left( \mathbf{I}-\mathbf{\Psi }{{\mathbf{\Psi }}^{*}} \right)\mathbf{\alpha } \right\|_{2}^{2}$ in a balanced model is $1/\gamma$ \cite{ 2017_MRM_bFISTA}, there are still strong demands for analyzing the convergence of pFISTA solving analysis models, or of FISTA solving the specific balanced model as they are still open questions. In other words, we do not know explicitly in advance which $\gamma$ can guarantee the algorithm to converge. Liu et al. \cite{2016_TMI_pFISTA} have proved that under the condition $\gamma =1$, pFISTA for single-coil MRI reconstruction is guaranteed to converge. Nevertheless, if the same setting, $\gamma =1$, is used in pFISTA-SENSE and pFISTA-SPIRiT, the algorithms may not converge (Fig. \ref{fig_RelatedWork}). This is because the sensitivity map or convolution kernel would affect the convergence property of pFISTA-parallel. We observed in experiments that a relatively large $\gamma$ leads to the divergence of pFISTA-parallel while a far smaller one results in the slow convergence of the algorithm (Fig. \ref{fig_RelatedWork}). Furthermore, the range of $\gamma$, allowing the algorithm to converge, varies under different tested data. Therefore, we aim to offer an explicit rule about how to choose a proper $\gamma$ of pFISTA-parallel to hold a fast convergence speed and promising results.

%----------------------------------------------------------------------
%------------------------ Convergence Analysis ------------------------
%----------------------------------------------------------------------

\section{Convergence Analysis} \label{Section:convergence}
In this section, we prove the convergence of pFISTA-parallel.

We present the analysis model of the parallel MRI reconstruction in a unified formula shown in Eq. \eqref{(pFISTA-parallel model)} in which the system matrix $\mathbf{A}$ has its explicit form $\mathbf{A}={\tilde{\mathbf{U}}}{\tilde{\mathbf{F}}}\mathbf{C}$ if the model is SENSE-based, and ${\bf{A}} = \left[ {\begin{array}{*{20}{c}}
{\tilde{\mathbf{U}}}{\tilde{\mathbf{F}}}&- \sqrt {{\lambda _1}} \left( {{\mathbf{W}} - {\mathbf{I}}} \right) \end{array}} \right]^{T}$ if the model is SPIRiT-based. According to \cite{2009_SIAM_FISTA,2016_TMI_pFISTA}, let $\left\{ {{\mathbf{d}}^{\left( k \right)}} \right\}$ be generated by pFISTA-parallel, and if the step size satisfies
\begin{equation}\label{(pFISTA-parallel convergence condition)}
  \gamma  \le \frac{1}{{L\left( \gamma  \right)}},
\end{equation}
and $\mathbf{\Psi }$ is a tight frame, the sequence $\left\{ {{\bm{\alpha }}^{\left( k \right)}} \right\}=\left\{ \mathbf{\Psi }{{\mathbf{d}}^{\left( k \right)}} \right\}$ converges to a solution of
\begin{equation}\label{(pFISTA-parallel convergence model)}
  \mathop {\min }\limits_{\bm{\alpha }} \lambda {\left\| {\bm{\alpha }} \right\|_1} + \frac{1}{2}\left\| {{\mathbf{y}} - {\mathbf{A}}{{\mathbf{\Psi }}^*}{\bm{\alpha }}} \right\|_2^2 + \frac{1}{{2\gamma }}\left\| {\left( {{\mathbf{I}} - {\mathbf{\Psi }}{{\mathbf{\Psi }}^*}} \right){\bm{\alpha }}} \right\|_2^2,
\end{equation}
with the speed
\begin{equation}\label{(pFISTA-parallel convergence speed)}
  F\left( {{{\bm{\alpha }}^{\left( k \right)}}} \right) - F\left( {{\bar{\bm \alpha }}} \right) \le \frac{2}{{\gamma {{\left( {k + 1} \right)}^2}}}{\left\| {{{\bm{\alpha }}^{\left( k \right)}} - {\bar{\bm \alpha }}} \right\|^2},
\end{equation}
where $\bar{\bm{\alpha }}$ is a solution of \eqref{(pFISTA-parallel convergence model)} and $F\left( \cdot  \right)$ is the objective function in \eqref{(pFISTA-parallel convergence model)} and $L$ is the Lipschitz constant for the gradient term.

Let us denote
\begin{equation}
\begin{aligned}
  & g\left( \bm{\alpha } \right)=\lambda {{\left\| \bm{\alpha} \right\|}_{1}}, \\ 
 & f\left( \bm{\alpha } \right)=\frac{1}{2\gamma }\left\| \left( \mathbf{I}-\mathbf{\Psi }{{\mathbf{\Psi }}^{*}} \right)\bm{\alpha } \right\|_{2}^{2}+\frac{1}{2}\left\| \mathbf{y}-\mathbf{A}{{\mathbf{\Psi }}^{*}}\bm{\alpha } \right\|_{2}^{2}. \\ 
\end{aligned}
\end{equation}
Then the Lipschitz constant is
\begin{equation}
	L\left( \gamma  \right)=L\left( \nabla f \right)={{\left\| \frac{1}{\gamma }\left( \mathbf{I}-\mathbf{\Psi }{{\mathbf{\Psi }}^{*}} \right)+\mathbf{\Psi }{{\mathbf{A}}^{H}}\mathbf{A}{{\mathbf{\Psi }}^{*}} \right\|}_{2}}.
\end{equation}
Let
\begin{equation}
	\mathbf{B}=\mathbf{\Psi }{{\mathbf{A}}^{H}}\mathbf{A}{{\mathbf{\Psi }}^{*}}-\frac{1}{\gamma }\mathbf{\Psi }{{\mathbf{\Psi }}^{*}}
\end{equation}
and $\mathbf{B}$ is a Hermitian matrix, then the matrix $\mathbf{B}+\frac{1}{\gamma }\mathbf{I}$ is also a Hermitian matrix. Therefore, we have
\begin{equation}
\begin{aligned}
	L\left( \gamma  \right)=&{{\left\| \mathbf{B}+\frac{1}{\gamma }\mathbf{I} \right\|}_{2}}=\underset{i}{\mathop{\max }}\,\left| {{e}_{i}}\left( \mathbf{B}+\frac{1}{\gamma }\mathbf{I} \right) \right|\\
  =&\underset{i}{\mathop{\max }}\,\left( \left| {{e}_{i}}\left( \mathbf{B} \right)+\frac{1}{\gamma } \right| \right),\\
\end{aligned}
\end{equation}
where ${{e}_{i}}\left( \cdot  \right)$ denotes the $i^{th}$ eigenvalue of matrix. Therefore, the key point is to analyze the eigenvalue of matrix $\mathbf{B}$.
Suppose $\mathbf{z}$ is an eigenvector of $\mathbf{B}$ corresponding to the eigenvalue $\beta $, by using the tight frame property, we have
\begin{equation}
	\begin{aligned}
  & \mathbf{Bz}=\left( \mathbf{\Psi }{{\mathbf{A}}^{H}}\mathbf{A}{{\mathbf{\Psi }}^{*}}-\frac{1}{\gamma }\mathbf{\Psi }{{\mathbf{\Psi }}^{*}} \right)\mathbf{z}=\beta \mathbf{z} \\ 
 & \Rightarrow \left( {{\mathbf{A}}^{H}}\mathbf{A}-\frac{1}{\gamma }\mathbf{I} \right)\left( {{\mathbf{\Psi }}^{*}}\mathbf{z} \right)=\beta \left( {{\mathbf{\Psi }}^{*}}\mathbf{z} \right), \\ 
\end{aligned}
\end{equation}
which indicates that all non-zero eigenvalues of $\mathbf{B}$ satisfy
\begin{equation}
	{{e}_{i}}\left( \mathbf{B} \right)\in \left\{ {{e}_{i}}\left( {{\mathbf{A}}^{H}}\mathbf{A}-\frac{1}{\gamma }\mathbf{I} \right) \right\}=\left\{ {{e}_{i}}\left( {{\mathbf{A}}^{H}}\mathbf{A} \right)-\frac{1}{\gamma } \right\}.
\end{equation}

Due to the redundancy, there exists $\bm{\alpha }\ne \mathbf{0}$ such that ${{\mathbf{\Psi }}^{*}}\bm{\alpha }=\mathbf{0}$. Thus, there are zero eigenvalues of $\mathbf{B}$:
\begin{equation}
  {{e}_{i}}\left( \mathbf{B} \right) \in \left \{ 0, \ e_{i} \left( {{\mathbf{A}}^{H}}\mathbf{A} \right) - \frac{1}{\gamma} \right \}.
\end{equation}
Therefore,
\begin{equation}
	L\left( \gamma  \right) \! = \! \underset{i}{\mathop{\max }} \! \left( \left| {{e}_{i}}\left( \mathbf{B} \right) \! + \! \frac{1}{\gamma } \right| \right) \! = \! \underset{i}{\mathop{\max }} \! \left\{ \frac{1}{\gamma },\left| e_{i} \left( {{\mathbf{A}}^{H}}\mathbf{A} \right) \right| \right\}.
\end{equation}

Now we are going to analyze the largest eigenvalue of ${{\mathbf{A}}^{H}}\mathbf{A}$ in different reconstruction problems. In the following, we will explicitly discuss the convergence of pFISTA-SENSE and pFISTA-SPIRiT.

\subsection{Convergence of pFISTA-SENSE}
In this section, we provide sufficient conditions for the convergence of pFISTA-SENSE in the form of a theorem.
\newtheorem{Theorem}{Theorem}
\begin{Theorem}\label{Theorem_SENSE}
Let $\left\{ \mathbf{x}^{\left(k \right)} \right\}$ be generated by pFISTA-SENSE, and if the sensitivity maps satisfies
\begin{equation}\label{(CSM)}
  {{\mathbf{C}}^{H}}\mathbf{C} = \mathbf{I},
\end{equation}
the step size satisfies
\begin{equation}\label{(pFSITA-SENSE convergence condition)}
  \gamma  \le 1,
\end{equation}
and $\mathbf{\Psi }$ is a tight frame, the sequence $\left\{ {{\bm{\alpha }}^{\left( k \right)}} \right\}=\left\{ \mathbf{\Psi }{{\mathbf{x}}_{c}^{\left( k \right)}} \right\}$ converges to a solution of
\begin{equation}\label{(pFISTA-SENSE convergence model)}
% \begin{aligned}
 \mathop {\min }\limits_{\bm{\alpha }} \lambda {\left\| {\bm{\alpha }} \right\|_1} \! + \! \frac{1}{2}\left\| {{\mathbf{y}} \! - \! \tilde{\mathbf{U}}\tilde{\mathbf{F}}\mathbf{C}{{\mathbf{\Psi }}^*}{\bm{\alpha }}} \right\|_2^2 \! + \! \frac{1}{{2\gamma }}\left\| {\left( {{\mathbf{I}} \! - \! {\mathbf{\Psi }}{{\mathbf{\Psi }}^*}} \right){\bm{\alpha }}} \right\|_2^2.
% \end{aligned}
\end{equation}
\end{Theorem}

\begin{proof}

In pFISTA-SENSE, we have $\mathbf{A}=\tilde{\mathbf{U}}\tilde{\mathbf{F}}\mathbf{C}$, thus,
\begin{equation}\label{(pFISTA-SENSE A*A-1)}
  {{\mathbf{A}}^H}{\mathbf{A}} = {{\mathbf{C}}^H}{{\tilde{\mathbf F}}^H}{{\tilde{\mathbf U}}^T}{\tilde{\mathbf{U}}\tilde{\mathbf{F}}\mathbf{C}}.
\end{equation}

Let $\mathbf{Q}={{\mathbf{F}}^{H}}{{\mathbf{U}}^{T}}\mathbf{UF}$, the matrix ${{\mathbf{C}}_j^H{\mathbf{Q}}{{\mathbf{C}}_j}} $ is a Hermitian matrix. For a Hermitian matrix, the largest eigenvalue is equal to the $\ell_2$ norm. In addition, notice that matrix $\ell_2$ norm satisfies triangle inequality and consistency property \cite{2000_Matrix_Analysis}, we can find the upper bound of the largest eigenvalue of the matrix ${{\mathbf{C}}^{H}}\mathbf{QC}$:
\begin{equation}\label{(pFISTA-SENSE C-1)}
  % \begin{aligned}
   \underset{i}{\mathop{\max }}\,{{e}_{i}}\left( {{\mathbf{C}}^{H}}\mathbf{QC} \right) \! = \! {{\left\| {{\mathbf{C}}^{H}}\mathbf{QC} \right\|}_{2}} \! \le \! {{\left\| {{\mathbf{C}}^{H}} \right\|}_{2}}{{\left\| \mathbf{Q} \right\|}_{2}}{{\left\| \mathbf{C} \right\|}_{2}}.
% \end{aligned}
\end{equation}
Here, the matrix $\mathbf{F}$ is a unitary matrix, according to the unitary invariant of $\ell_2$ norm, we have
\begin{equation}\label{(Q-1)}
  {\left\| {\mathbf{Q}} \right\|_2} = {\left\| {{{\mathbf{F}}^H}{{\mathbf{U}}^T}{\mathbf{UF}}} \right\|_2} = {\left\| {{{\mathbf{U}}^T}{\mathbf{U}}} \right\|_2}.
\end{equation}
And ${{\mathbf{U}}^{T}}\mathbf{U}$ is a diagonal matrix with the diagonal elements $0$ or $1$, indicating that
\begin{equation}\label{(Q-2)}
  {\left\| {\mathbf{Q}} \right\|_2} = 1.
\end{equation}

With Eq. \eqref{(Q-2)}, we can further simplify Eq. \eqref{(pFISTA-SENSE C-1)}
\begin{equation}\label{(pFISTA-SENSE C-2)}
  % \begin{aligned}
   \underset{i}{\mathop{\max }}\,{{e}_{i}} \! \left( {{\mathbf{C}}^{H}}\mathbf{QC} \right) \! \le \! {{\left\| {{\mathbf{C}}^{H}} \right\|}_{2}}{{\left\| \mathbf{Q} \right\|}_{2}}{{\left\| \mathbf{C} \right\|}_{2}} \! = \! {{\left\| {{\mathbf{C}}^{H}} \right\|}_{2}}{{\left\| \mathbf{C} \right\|}_{2}}.
 % \end{aligned}
\end{equation}

Here, if $\mathbf{C}$ is normalized such that ${{\mathbf{C}}^{H}}\mathbf{C}=\mathbf{I}$, which is quite common in practice, we can further obtain
\begin{equation}\label{(pFISTA-SENSE C-3)}
{{\left\| \mathbf{C} \right\|}_{2}}\text{=}\sqrt{\underset{i}{\mathop{\text{max}}}\,{{e}_{i}}\left( {{\mathbf{C}}^{H}}\mathbf{C} \right)}=\sqrt{\underset{i}{\mathop{\text{max}}}\,{{e}_{i}}\left( \mathbf{I} \right)}=1,
\end{equation}
and at the same time
\begin{equation}\label{(pFISTA-SENSE C-4)}
{{\left\| {{\mathbf{C}}^{H}} \right\|}_{2}}={{\left\| \mathbf{C} \right\|}_{2}}=1.
\end{equation}
Finally, we have 
\begin{equation}\label{(pFISTA-SENSE C-5)}
\underset{i}{\mathop{\max }}\,{{e}_{i}}\left( {{\mathbf{C}}^{H}}\mathbf{QC} \right)\le {{\left\| {{\mathbf{C}}^{H}} \right\|}_{2}}{{\left\| \mathbf{Q} \right\|}_{2}}{{\left\| \mathbf{C} \right\|}_{2}}=1.
\end{equation}

Therefore,
\begin{equation}\label{(pFISTA-SENSE convergence)}
%\begin{medsize}
  \begin{array}{l}
    L\left( \gamma  \right) = \mathop {\max }\limits_i \left\{ {\frac{1}{\gamma },\left| {{e_i}\left( {{\mathbf{C}}^H{\mathbf{Q}}{{\mathbf{C}}}} \right)} \right|} \right\} = \frac{1}{\gamma }, \ 0 < \gamma  \le 1,\\
    L\left( \gamma  \right) = \mathop {\max }\limits_i \left\{ {\frac{1}{\gamma },\left| {{e_i}\left( {{\mathbf{C}}^H{\mathbf{Q}}{{\mathbf{C}}}}  \right)} \right|} \right\} = 1, \ \gamma  > 1.
  \end{array}
%\end{medsize}
\end{equation}

The Eq. \eqref{(pFISTA-SENSE convergence)} means that, when $0<\gamma \le 1$, one has $L\left( \gamma  \right)=1/\gamma$, which satisfies the convergence condition of pFISTA in Eq. \eqref{(pFISTA-parallel convergence condition)}; whereas when $\gamma >1$, then $L\left( \gamma  \right)=1>1/\gamma$, which does not satisfy the convergence condition of pFISTA. In summary, when $0<\gamma \le 1$, the pFISTA-SENSE is guaranteed to converge.
\end{proof}

\subsection{Convergence of pFISTA-SPIRiT}
In this section, we provide sufficient conditions for the convergence of pFISTA-SPIRiT in the form of a theorem.
\begin{Theorem}\label{Theorem_SPIRiT}
Let $\left\{ \mathbf{x}^{\left(k \right)} \right\}$ be generated by pFISTA-SPIRiT, and if the step size satisfies
% \begin{equation}\label{(pFSITA-SPIRiT theorem)}
% \begin{medsize}
% \begin{array}{c}
%   {\gamma}  \le \frac{1}{c},\\
%   c=1+{{\lambda }_{1}}\left( {{\left\| {{\mathbf{Z}}_{diag,0}} \right\|}_{2}}+2\times \sum\limits_{i=1}^{z}{{{\left\| {{\mathbf{Z}}_{diag,i}} \right\|}_{2}}}+\sum\limits_{i=z+1}^{J-1}{{{\left\| {{\mathbf{Z}}_{diag,i}} \right\|}_{2}}} \right),
%   \end{array}
%   \end{medsize}
% \end{equation}
\begin{equation}\label{(pFSITA-SPIRiT theorem)}
  {\gamma}  \le  \frac{1}{c}, \quad c =\sum\limits_{i=-z}^{z}{{{\left\| {{\mathbf{Z}}_{diag,i}} \right\|}_{2}}}+\sum\limits_{i=z+1}^{J-1}{{{\left\| {{\mathbf{Z}}_{diag,i}} \right\|}_{2}}},
\end{equation}
and $\mathbf{\Psi }$ is a tight frame, the sequence $\left\{ {{\bm{\alpha }}^{\left( k \right)}} \right\}=\left\{ \mathbf{\Psi }{{\mathbf{x}}^{\left( k \right)}} \right\}$ converges to a solution of
\begin{equation}\label{(pFISTA-SPIRiT convergence model)}
\begin{aligned}
 \mathop {\min }\limits_{\bm{\alpha }} &\lambda {\left\| {\bm{\alpha }} \right\|_1} + \frac{1}{{2\gamma }}\left\| {\left( {{\mathbf{I}} - {\mathbf{\Psi }}{{\mathbf{\Psi }}^*}} \right){\bm{\alpha }}} \right\|_2^2 \\
&+\frac{1}{2}\left\| {{\mathbf{y}} - \left[ {\begin{array}{*{20}{c}}
        {{\tilde{\mathbf U}} \tilde{\mathbf F}}\\
        { - \sqrt {{\lambda _1}} \left( {{\mathbf{W}} - {\mathbf{I}}} \right)}
        \end{array}} \right]{{{\mathbf{\Psi }}^*}{\bm{\alpha }}}} \right\|_2^2.
\end{aligned}
\end{equation}
\end{Theorem}

\begin{proof}

In pFISTA-SPIRiT, we have
\begin{equation}\label{(pFISTA-SPIRiT A*A)}
\begin{medsize}
\begin{aligned}
    {{\mathbf{A}}^H}{\mathbf{A}} &= \left[ {\begin{array}{*{20}{c}}
    {{{{\tilde{\mathbf F}}}^H} {{{\tilde{\mathbf U}}}^T}}&{ - \sqrt {{\lambda _1}} {{\left( {{\mathbf{W}} - {\mathbf{I}}} \right)}^H}}
    \end{array}} \right]\left[ {\begin{array}{*{20}{c}}
    {{\tilde{\mathbf U}} {\tilde{\mathbf F}}}\\
    { - \sqrt {{\lambda _1}} \left( {{\mathbf{W}} - {\mathbf{I}}} \right)}
    \end{array}} \right]\\
    &= \left( {{{\tilde{\mathbf F}}}^H} {{{{\tilde{\mathbf U}}}^T}{\tilde{\mathbf U}}{\tilde{\mathbf F}} + {\lambda _1}{{\left( {{\mathbf{W}} - {\mathbf{I}}} \right)}^H}\left( {{\mathbf{W}} - {\mathbf{I}}} \right)} \right).
  \end{aligned}
\end{medsize}
\end{equation}
Since $ {{\mathbf{A}}^H}{\mathbf{A}} $ is a Hermitian matrix, the maximum eigenvalue of $ {{\mathbf{A}}^H}{\mathbf{A}} $ equals its $\ell_2$ norm. And according to the linearity and triangle inequality of matrix norm \cite{2000_Matrix_Analysis}, we have
\begin{equation}\label{(pFISTA-SPIRiT A*A-2)}
\begin{aligned}
& \;\;\;\; {{e}_{i}}\left( {{\mathbf{A}}^H}{\mathbf{A}} \right) \\ 
 & ={{\left\| {{{\tilde{\mathbf F}}}^H} {{{\tilde{\mathbf U}}}^T}{\tilde{\mathbf U}}{\tilde{\mathbf F}} + {{\lambda }_{1}}{{\left( \mathbf{W}-\mathbf{I} \right)}^{H}}\left( \mathbf{W}-\mathbf{I} \right) \right\|}_{2}} \\ 
 & \le {{\left\| {{{\tilde{\mathbf F}}}^H} {{{\tilde{\mathbf U}}}^T}{\tilde{\mathbf U}}{\tilde{\mathbf F}} \right\|}_{2}}+{{\lambda }_{1}}{{\left\| {{\left( \mathbf{W}-\mathbf{I} \right)}^{H}}\left( \mathbf{W}-\mathbf{I} \right) \right\|}_{2}}.
 \end{aligned}
\end{equation}
As ${\tilde{\mathbf F}}$ is also a unitary matrix, the same as the proof in the previous subsection, we can easily derive that $\left\| {{{\tilde{\mathbf F}}}^H} {{{\tilde{\mathbf U}}}^T}{\tilde{\mathbf U}}{\tilde{\mathbf F}} \right\|_2 = 1$. Thus, we can rewrite \eqref{(pFISTA-SPIRiT A*A-2)} as
\begin{equation}
\begin{aligned}
& \;\;\;\;{{e}_{i}} \left( {\mathbf{A}}^H {\mathbf A} \right)\\
& \le {{\left\| {{{\tilde{\mathbf F}}}^H} {{{\tilde{\mathbf U}}}^T}{\tilde{\mathbf U}}{\tilde{\mathbf F}} \right\|}_{2}}+{{\lambda }_{1}}{{\left\| {{\left( \mathbf{W}-\mathbf{I} \right)}^{H}}\left( \mathbf{W}-\mathbf{I} \right) \right\|}_{2}}\\
& =1+{{\lambda }_{1}}{{\left\| {{\left( \mathbf{W}-\mathbf{I} \right)}^{H}}{\left( \mathbf{W}-\mathbf{I} \right)} \right\|}_{2}}.
\end{aligned}
\end{equation}

% Table generated by Excel2LaTeX from sheet 'Shee$T_{1}$'
\begin{table*}[!htp]
  \centering
  \caption{Detailed MRI scanning parameters for the data used in this work.}
    \begin{tabular}{cccccccc}
    \toprule
    Data & Scanner & Sequence & Number of coils & Matrix size & TR/TE (ms) & FOV (mm$^{2}$) & Slice thickness (mm) \\
    \midrule
    Fig. 3 (a) & 1.5T Philips & $T_{1}$-weighted fast-field-echo & 8     & 256*256 & 1700/390 & 230*230 & 5 \\
    Fig. 3 (b) & 3T GE & $T_{1}$- weighted SPGR & 12    & 256*256 & 400/9 & 240*240 & 6 \\
    Fig. 3 (c) & 3T Siemens & $T_{2}$-weighted turbo spin echo & 32    & 256*256 & 6100/99 & 220*220 & 3 \\
    \bottomrule
    \end{tabular}%
  \label{Data}%
\end{table*}%

Indeed, once the kernels have been estimated using ACS, the matrix $\mathbf{W}$ is determined, which indicates that the maximum eigenvalue of ${\mathbf{A}^H} \mathbf{A}$ can be obtained. However, the computation of the system matrix's $\ell_2$ norm poses as a challenging task due to the huge dimensionality of the matrix, for example, as for an 8-coil $256\times 256$ image, the size of $\mathbf{W}$ reaches $524288\times 524288$. Therefore, we further relax bound so as to calculate it efficiently.

Let $\mathbf{D}=\mathbf{W}-\mathbf{I}$, that is
\begin{equation}
\begin{medsize}
\mathbf{D} \! = \! \left[\begin{matrix}
   {{\mathbf{D}}_{1,1}} & {{\mathbf{D}}_{1,2}} & \cdots  & {{\mathbf{D}}_{1,J}}  \\
   {{\mathbf{D}}_{2,1}} & {{\mathbf{D}}_{2,2}} & \cdots  & {{\mathbf{D}}_{2,J}}  \\
   \vdots  & \vdots  & \ddots  & \vdots   \\
   {{\mathbf{D}}_{J,1}} & {{\mathbf{D}}_{J,2}} & \cdots  & {{\mathbf{D}}_{J,J}}  \\
\end{matrix} \right]\ \! \text{with}\;{{\mathbf{D}}_{i,j}} \! = \! \left\{
\begin{aligned}
  & {{\mathbf{W}}_{i,j}} \! - \! \mathbf{I},\;i=j, \\ 
 & {{\mathbf{W}}_{i,j}},\; \; \; \; \; \; i\ne j.
\end{aligned}
\right.
\end{medsize}
\end{equation}
And denote $\mathbf{Z}={{\left( \mathbf{W}-\mathbf{I} \right)}^{H}} {\left( \mathbf{W}-\mathbf{I} \right)} $, we have 

\begin{equation}
\begin{medsize}
\mathbf{Z} = \left[ \begin{matrix}
   {{\mathbf{Z}}_{1,1}} & {{\mathbf{Z}}_{1,2}} & \cdots  & {{\mathbf{Z}}_{1,J}}  \\
   {{\mathbf{Z}}_{2,1}} & {{\mathbf{Z}}_{2,2}} & \cdots  & {{\mathbf{Z}}_{2,J}}  \\
   \vdots  & \vdots  & \ddots  & \vdots   \\
   {{\mathbf{Z}}_{J,1}} & {{\mathbf{Z}}_{J,2}} & \cdots  & {{\mathbf{Z}}_{J,J}}  \\
\end{matrix} \right]\ \text{with} \; {{\mathbf{Z}}_{i,j}} = \sum\limits_{m=1}^{J}{\mathbf{D}_{m,i}^{H}{{\mathbf{D}}_{m,j}}}
\end{medsize}
\end{equation}
and each ${\mathbf{Z}}_{i,j}$ is a diagonal matrix. Then we can express $\mathbf{Z}$ as a superposition of block matrices:

\begin{equation}\label{(pFISTA-SPIRiT-Z-1)}
\begin{medsize}
\begin{aligned}
  \mathbf{Z} = &
    \underbrace{\left[ \begin{matrix}
   {{\mathbf{Z}}_{1,1}} & {} & {}  \\
   {} & \ddots  & {}  \\
   {} & {} & {{\mathbf{Z}}_{J,J}}  \\
\end{matrix} \right] }_{\mathbf{Z}_{diag,0}} + \\ 
 & \underbrace{\left[ \begin{matrix}
   {} & {{\mathbf{Z}}_{1,2}} & {} & {}  \\
   {} & {} & \ddots  & {}  \\
   {} & {} & {} & {{\mathbf{Z}}_{J-1,J}}  \\
   {} & {} & {} & {}  \\
\end{matrix} \right] }_{\mathbf{Z}_{diag,1}} +\cdots + \underbrace{\left[ \begin{matrix}
   {} & {} & {} & {{\mathbf{Z}}_{1,J}}  \\
   {} & {} & {} & {}  \\
   {} & {} & {} & {}  \\
   {} & {} & {} & {}  \\
\end{matrix} \right] }_{\mathbf{Z}_{diag,J-1}} + \\ 
 & \underbrace{ \left[ \begin{matrix}
   {} & {} & {} & {}  \\
   {{\mathbf{Z}}_{2,1}} & {} & {} & {}  \\
   {} & \ddots  & {} & {}  \\
   {} & {} & {{\mathbf{Z}}_{J,J-1}} & {}  \\
\end{matrix} \right] }_{\mathbf{Z}_{diag,-1}} +\cdots + \underbrace{\left[ \begin{matrix}
   {} & {} & {} & {}  \\
   {} & {} & {} & {}  \\
   {} & {} & {} & {}  \\
   {{\mathbf{Z}}_{J,1}} & {} & {} & {}  \\
\end{matrix} \right] }_{\mathbf{Z}_{diag,-(J-1)}}
\end{aligned}
\end{medsize}
\end{equation}

Let ${\mathbf{Z}}_{diag,i}$ denotes the $i$-block matrix. We can rewrite Eq. \eqref{(pFISTA-SPIRiT-Z-1)} as:
\begin{equation}
\mathbf{Z}=\sum\limits_{i=-\left( J-1 \right)}^{J-1}{{{\mathbf{Z}}_{diag,i}}}.
\end{equation}

\begin{figure}[!hb]
\setlength{\abovecaptionskip}{0.cm}
\setlength{\belowcaptionskip}{-0.cm}
\centering
\includegraphics[width=2.2in]{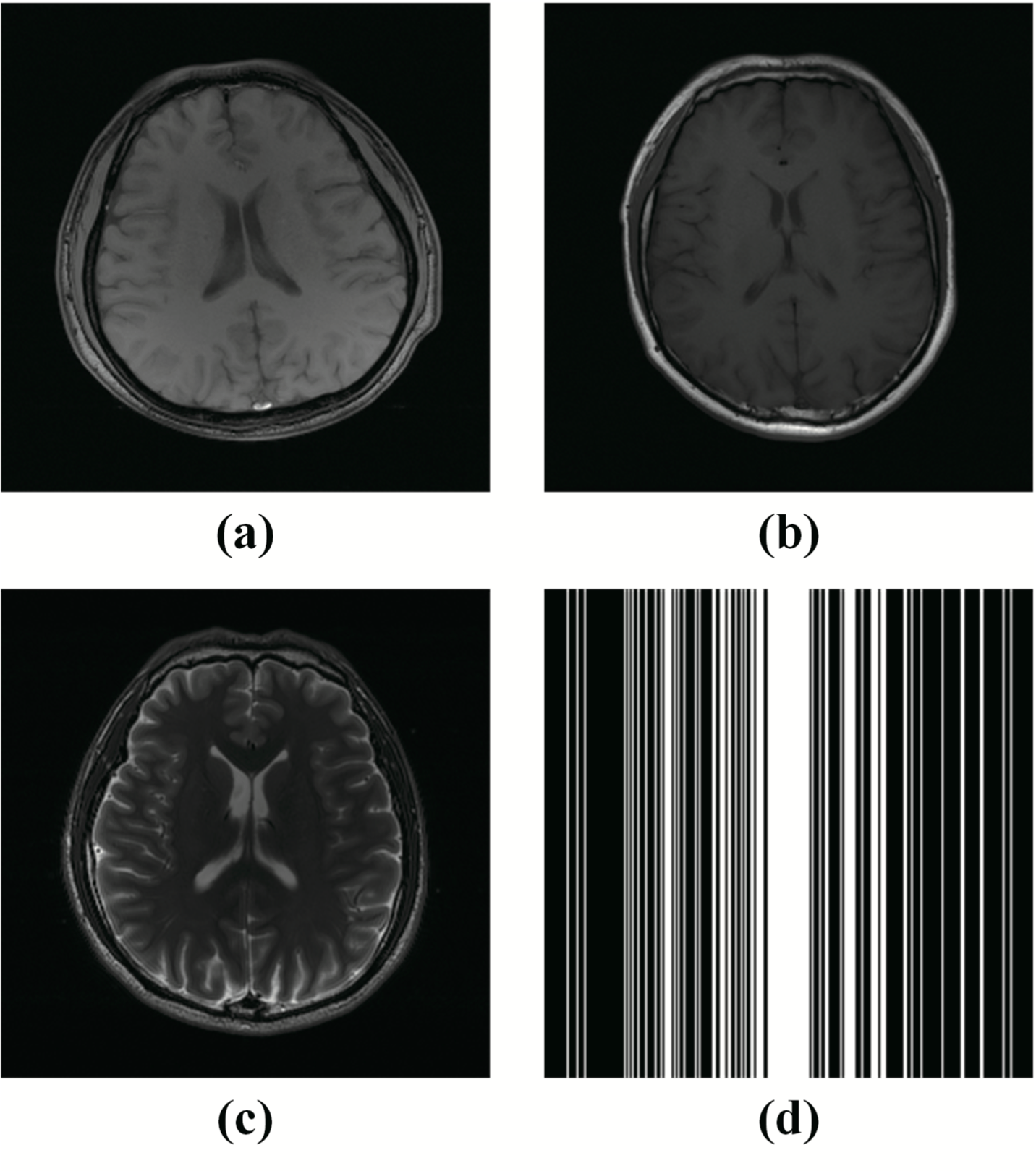}
\caption{Experimental dataset. (a-c) Three different brain images; (d) the Cartesian sampling pattern of sampling rate 0.34.}
\label{fig_Dataset}
\end{figure}

\begin{figure*}[!htb]
\setlength{\abovecaptionskip}{0.cm}
\setlength{\belowcaptionskip}{-0.cm}
\centering
\includegraphics[width=5.6in]{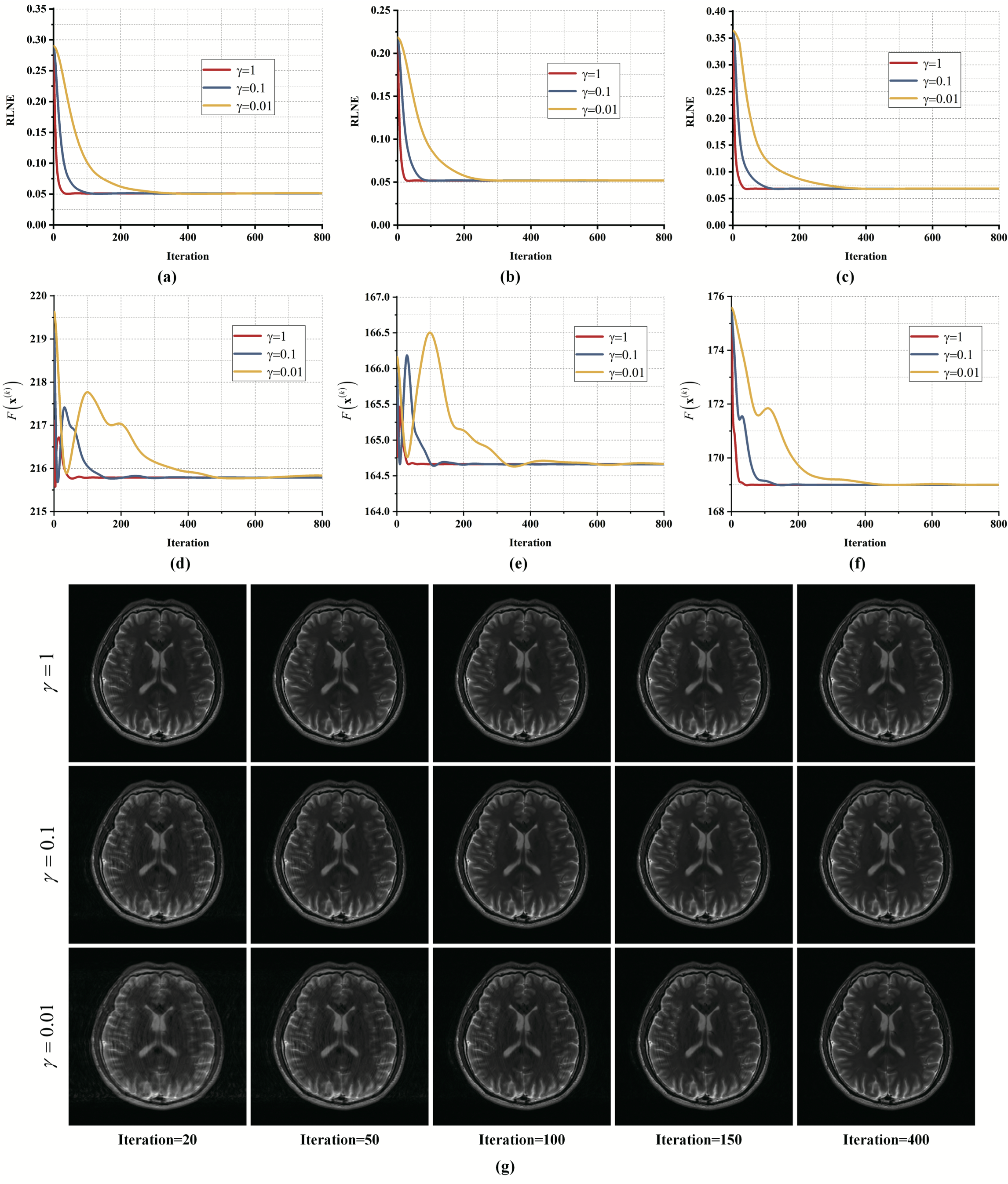}
\caption{Reconstructions of three different brain images by pFISTA-SENSE with different step size $\gamma$. (a-c) are the RLNEs of pFISTA-SENSE using 8, 12 and 32-coil data shown in Fig. \ref{fig_Dataset}, respectively. (d-f) are the function values of pFISTA-SENSE using 8, 12 and 32-coil data shown in Fig. \ref{fig_Dataset}, respectively. (g) is the reconstructions of pFISTA-SENSE at different iteration in 32-coil data. All experiments used the same sampling pattern depicted in Fig. \ref{fig_Dataset}.}
\label{fig_results_pFISTA-SENSE}
\end{figure*}

Then, 
\begin{equation}\label{(pFISTA-SPIRiT-Z)}
{{\left\| \mathbf{Z} \right\|}_{2}}={{\left\| \sum\limits_{i=-\left( J-1 \right)}^{J-1}{{{\mathbf{Z}}_{diag,i}}} \right\|}_{2}}.
\end{equation}

We want to point out that
\begin{equation}\label{(pFISTA-SPIRiT-P2)}
{{\left\| {{\mathbf{Z}}_{diag,i}}+{{\mathbf{Z}}_{diag,-i}} \right\|}_{2}}={{\left\| {{\mathbf{Z}}_{diag,i}} \right\|}_{2}},\ i=1,\cdots ,z,
\end{equation}
where $z$ denotes the biggest integer no more than $J/2$. Detailed proof of Eq. \eqref{(pFISTA-SPIRiT-P2)} can be found in Supplementary Material.
With the help of Eq. \eqref{(pFISTA-SPIRiT-P2)}, we can rewrite the Eq. \eqref{(pFISTA-SPIRiT-Z)} as:
% \begin{equation}
% \begin{medsize}
% \begin{aligned}
%    {{\left\| \mathbf{Z} \right\|}_{2}} & ={{\left\| \sum\limits_{i=-\left( J+1 \right)}^{J-1}{{{\mathbf{Z}}_{diag,i}}} \right\|}_{2}} \\ 
%  & ={{\left\| {{\mathbf{Z}}_{diag,0}}+\sum\limits_{i=1}^{z}{{{\mathbf{Z}}_{diag,i}}}+\sum\limits_{i=-z}^{-1}{{{\mathbf{Z}}_{diag,i}}}+\sum\limits_{i=z+1}^{J-1}{\left( {{\mathbf{Z}}_{diag,i}}+{{\mathbf{Z}}_{diag,-i}} \right)} \right\|}_{2}} \\ 
%  & \le {{\left\| {{\mathbf{Z}}_{diag,0}} \right\|}_{2}}+\sum\limits_{i=1}^{z}{{{\left\| {{\mathbf{Z}}_{diag,i}} \right\|}_{2}}}+\sum\limits_{i=-z}^{-1}{{{\left\| {{\mathbf{Z}}_{diag,i}} \right\|}_{2}}}\\
%  & \; \; \; \; \;  +\sum\limits_{i=z+1}^{J-1}{{{\left\| {{\mathbf{Z}}_{diag,i}}+{{\mathbf{Z}}_{diag,-i}} \right\|}_{2}}} \\ 
%  & ={{\left\| {{\mathbf{Z}}_{diag,0}} \right\|}_{2}}+2\times \sum\limits_{i=1}^{z}{{{\left\| {{\mathbf{Z}}_{diag,i}} \right\|}_{2}}}+\sum\limits_{i=z+1}^{J-1}{{{\left\| {{\mathbf{Z}}_{diag,i}} \right\|}_{2}}}.  
% \end{aligned}
% \end{medsize}
% \end{equation}
\begin{equation}
\begin{medsize}
\begin{aligned}
 {{\left\| \mathbf{Z} \right\|}_{2}}  = &{{\left\| \sum\limits_{i=-\left( J+1 \right)}^{J-1}{{{\mathbf{Z}}_{diag,i}}} \right\|}_{2}} \\ 
 = & {{\left\|\! \sum\limits_{i=-z}^{z}{{{\mathbf{Z}}_{diag,i}}} \! + \! \sum\limits_{i=z+1}^{J-1}{\left( {{\mathbf{Z}}_{diag,i}} \! + \! {{\mathbf{Z}}_{diag,-i}} \right)} \right\|}_{2}} \\ 
  \le & \sum\limits_{i=-z}^{z}{{{\left\| {{\mathbf{Z}}_{diag,i}} \right\|}_{2}}}+\sum\limits_{i=z+1}^{J-1}{{{\left\| {{\mathbf{Z}}_{diag,i}}+{{\mathbf{Z}}_{diag,-i}} \right\|}_{2}}} \\ 
 =& \sum\limits_{i=-z}^{z}{{{\left\| {{\mathbf{Z}}_{diag,i}} \right\|}_{2}}}+\sum\limits_{i=z+1}^{J-1}{{{\left\| {{\mathbf{Z}}_{diag,i}} \right\|}_{2}}}.  
\end{aligned}
\end{medsize}
\end{equation}

Therefore, the maximum eigenvalue of the system matrix can be estimated by:
\begin{equation}\label{(pFSITA-SPIRiT C-2)}
\begin{medsize}
  \begin{aligned}
  & {{e}_{i}}\left( {\mathbf{A}^H}{\mathbf A} \right) \\ 
 \le & {{\left\| {{{\tilde{\mathbf F}}}^H} {{{\tilde{\mathbf U}}}^T}{\tilde{\mathbf U}}{\tilde{\mathbf F}} \right\|}_{2}}+{{\lambda }_{1}}{{\left\| {{\left( \mathbf{W}-\mathbf{I} \right)}^{H}}\left( \mathbf{W}-\mathbf{I} \right) \right\|}_{2}} \\ 
 = & 1+{{\lambda }_{1}}{{\left\| {{\left( \mathbf{W}-\mathbf{I} \right)}^{H}}\left( \mathbf{W}-\mathbf{I} \right) \right\|}_{2}} \\ 
 \le & \sum\limits_{i=-z}^{z}{{{\left\| {{\mathbf{Z}}_{diag,i}} \right\|}_{2}}}+\sum\limits_{i=z+1}^{J-1}{{{\left\| {{\mathbf{Z}}_{diag,i}} \right\|}_{2}}}. \\
\end{aligned}
\end{medsize}
\end{equation}

Let $c=\sum\limits_{i=-z}^{z}{{{\left\| {{\mathbf{Z}}_{diag,i}} \right\|}_{2}}}+\sum\limits_{i=z+1}^{J-1}{{{\left\| {{\mathbf{Z}}_{diag,i}} \right\|}_{2}}}$, we have
\begin{equation}\label{(pFISTA-SPIRiT convergence)}
\begin{medsize}
  \begin{aligned}
    L\left( \gamma  \right) & = \mathop {\max }\limits_i \left\{ {\frac{1}{\gamma },\left| {{e_i}\left( \mathbf{A}^H \mathbf{A} \right)} \right|} \right\} = \frac{1}{\gamma },\quad 0 < \gamma  \le \frac{1}{c},\\
    L\left( \gamma  \right) &= \mathop {\max }\limits_i \left\{ {\frac{1}{\gamma },\left| {{e_i}\left( \mathbf{A}^H \mathbf{A} \right)} \right|} \right\} = c,\quad \;\; \gamma  > \frac{1}{c}.
  \end{aligned}
\end{medsize}
\end{equation}

The Eq. \eqref{(pFISTA-SPIRiT convergence)} means that, when $0<\gamma \le 1/c$, one has $L\left( \gamma  \right)=1/{\gamma }$, which satisfies the convergence condition of pFISTA; whereas when $\gamma >1/c$, then $L\left( \gamma  \right)=c>1/{\gamma }$, which does not satisfy the convergence condition of pFISTA. In summary, when $0<\gamma \le 1/c$, pFISTA-SPIRiT is guaranteed to converge.
\end{proof}

%----------------------------------------------------------------------
%------------------------------- Results ------------------------------
%----------------------------------------------------------------------
\begin{figure*}[!htb]
\setlength{\abovecaptionskip}{0.cm}
\setlength{\belowcaptionskip}{-0.cm}
\centering
\includegraphics[width=5.6in]{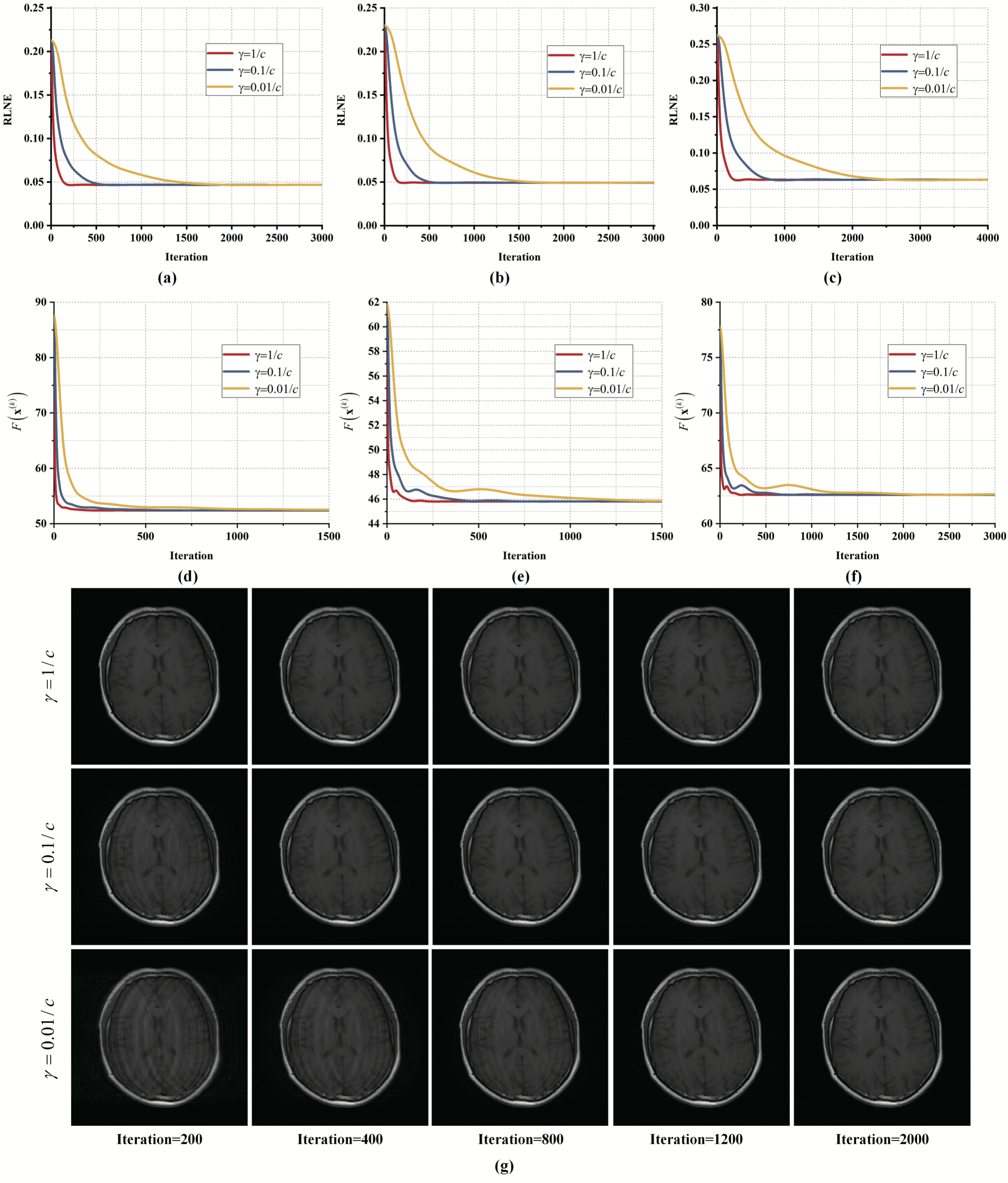}
\caption{Reconstructions of three different brain images by pFISTA-SPIRiT with different step size $\gamma$. (a-c) are the RLNEs of pFISTA-SPIRiT using 8, 12 and 32-coil data shown in Fig. \ref{fig_Dataset}, respectively. (d-f) are the function values of pFISTA-SPIRiT using 8, 12 and 32-coil data shown in Fig. \ref{fig_Dataset}, respectively. (g) is the reconstructions of pFISTA-SPIRiT at different iteration in 12-coil data. All experiments used the same sampling pattern depicted in Fig. \ref{fig_Dataset}.}
\label{fig_results_pFISTA-SPIRiT}
\end{figure*}

\section{Experimental Results}\label{Section:experimentalResults}
In this section, we first conducted experiments on multi-coils MRI brain images to assess the feasibility of the convergence criteria we derived. Then, we assess the gap between the proved sufficient condition and the hand-tuned optimal parameter and find that this gap leads to no distinct difference of convergence speeds between the sufficient condition and the hand-tuned optimal parameter. Besides, we made comparisons with approaches that allow the computation of the step size $\gamma$, such as backtracking and power iteration. Furthermore, we compared the reconstructions of pFISTA-parallel and other widely adopted algorithms - ADMM \cite{2016_TMI_pFISTA} and NLCG \cite{2007_MRM_Sparse_MRI}. The ADMM and NLCG software to solve SENSE and SPIRiT analysis reconstruction models were implemented by ourselves. Besides, comparisons with other FISTA algorithms were made, including MFISTA-FGP \cite{2009_TIP_FISTA_FGP} and MFISTA-VA \cite{2018_TCI_MFISTA_VA}. The codes of MFISTA-FGP and MFISTA-VA are shared on-line by Dr. Marcelo Zibetti \cite{MFISTA_codes}. Last, we discussed the convergence and results under other tight frames with different $\gamma$.

We adopted the objective-functional-based criteria $F ( \mathbf{x}^{(k)} ) $ to assess the convergence of algorithms. Here $F$ is the algorithm's function value, $\mathbf{x}^{(k)}$ is the solution of $k^{th}$ iteration. The function value of pFISTA is $F(\mathbf{x}) = \lambda {\left\| {\mathbf{x }} \right\|_1} + \frac{1}{2}\left\| {{\mathbf{y}} - {\mathbf{A}}{{\mathbf{\Psi }}^*}{\mathbf{x }}} \right\|_2^2 + \frac{1}{{2\gamma }}\left\| {\left( {{\mathbf{I}} - {\mathbf{\Psi }}{{\mathbf{\Psi }}^*}} \right){\mathbf{x }}} \right\|_2^2$.

Besides, relative $\ell_2$ norm error (RLNE) is also adopted to quantify the reconstruction performance. The RLNE is defined as
\begin{equation}\label{(RLNE)}
  {\rm{RLNE}} = \frac{{{{\left\| {{{\mathbf{x}}_{ref}} - {{\mathbf{x}}_{rec}}} \right\|}_2}}}{{{{\left\| {{{\mathbf{x}}_{ref}}} \right\|}_2}}},
\end{equation}
where ${{\mathbf{x}}_{ref}}$ denotes the vectorized reference image that is a square root of the sum of squares (SSOS) of the fully sampled image and ${{\mathbf{x}}_{rec}}$ the vectorized reconstructed image that is the SSOS image of pFISTA-SPIRiT reconstructed image and modular image of pFISTA-SENSE reconstructed image. We should point out that a lower RLNE, a higher consistency between the reference image and the reconstructed image.

Three multi-coil MRI datasets acquired from healthy volunteers are used in experiments. We list the detailed MRI scanning parameters in Table \ref{Data}.
For SENSE, the fully sampled $256 \times 64$ areas of the k-space center are used to calculate sensitivity maps \cite{2007_JSENSE}, and for SPIRiT, a fully sampled $256 \times 22$ areas for the Cartesian sampling pattern of sampling ratio $0.34$ are used to estimate the convolution kernels. The shift-invariant discrete wavelets transform (SIDWT) \cite{2011_SIDWT_Ying, 1995_SIDWT_Denosing, 2014_MRI_SIDWT}, if not mentioned otherwise, is adopted as the tight frame in experiments. In all experiments involving SIDWT, Daubechies wavelets with 4 decomposition levels are utilized. For pFISTA-SENSE, $\lambda ={{10}^{-3}}$ is set and for pFISTA-SPIRiT, we set $\lambda ={{10}^{-4}}$ and ${{\lambda }_{1}}=1$, and $5 \times 5$ SPIRiT kernel is used. All computation procedures run on a CentOS 7 computation server with two Intel Xeon CPUs of $3.5$ GHz and $112$ GB RAM.

\subsection{Main Results}
As mentioned above, once the parameter meets the condition $0<\gamma \le 1/c$, both the pFISTA-SENSE and pFISATA-SPIRiT converge. Thus, here we perform reconstructions by pFISTA-SENSE and pFISTA-SPIRiT with various $\gamma $ in the recommended range, respectively, to verify if the recommended $\gamma $ could enable the convergence of the algorithm.

As shown in Fig. \ref{fig_results_pFISTA-SENSE} (d-f), for three tested brain images, pFISTA-SENSE converges with the $\gamma $ ranged from $0.01/c$ to $1/c$. 
Moreover, the larger the $\gamma $, the faster the algorithm converges, this observation is consistent with the Eq. \eqref{(pFISTA-SENSE convergence)}. Notably, the upper bound of $\gamma$ is $1$ (here $c=1$), which, in other words, manifests that the number of coils does not relate to the convergence of pFISTA-SENSE. Besides, the behavior of RLNE also indicates the same phenomenon. A larger $\gamma$ indicates a faster speed to reach the final RLNE level. Particularly, $\gamma=1$ enables the fastest reconstruction. Please note that the faster the algorithm to reach the final RLNE level, the lesser the time is needed for reconstruction.

The intermediate reconstructed images manifest the convergence speeds of pFISTA-SENSE with various $\gamma$. The undersampling artifacts were quickly removed within $50$ iterations when with parameter $\gamma = 1$, and the algorithm produced a promising reconstructed image (Fig. \ref{fig_results_pFISTA-SENSE} (d)). As $\gamma$ decreased, the algorithm took more time to converge to a stage that yields satisfying results. For instance, when $\gamma=0.01$, the convergence criterion $F ( x^{(k)})$ still at a relatively high level even after $150$ iterations. The program eventually took about $400$ iterations to eliminate the obvious artifacts. In a word, the convergence criteria we provided can escort pFISTA-SENSE to achieve satisfying results of parallel imaging experiments. Furthermore, we would recommend using $\gamma=1$ for SENSE reconstructions.

In addition, we observe a similar phenomenon on pFISTA-SPIRiT experiments (Fig. \ref{fig_results_pFISTA-SPIRiT}). With $0<\gamma \le 1/c$, pFISTA-SPIRiT empirically converges but at different ratios. The larger the $\gamma $, the faster the algorithm converges. The fastest convergence speed is achieved when $\gamma =1/c$, thus we recommend $\gamma =1/c$ for pFISTA-SPIRiT experiments. The intermediate results of pFISTA-SPIRiT with monotonically decreasing $\gamma$ also reveal the increasing convergence rate as $\gamma$ rises (Fig. \ref{fig_results_pFISTA-SPIRiT}. The pFISTA-SPIRiT could be applied in multi-coils imaging experiments with guaranteed convergence if $0<\gamma \le 1/c$.

\begin{figure}[htb]
\setlength{\abovecaptionskip}{0.cm}
\setlength{\belowcaptionskip}{-0.cm}
\centering
\includegraphics[width=3.5in]{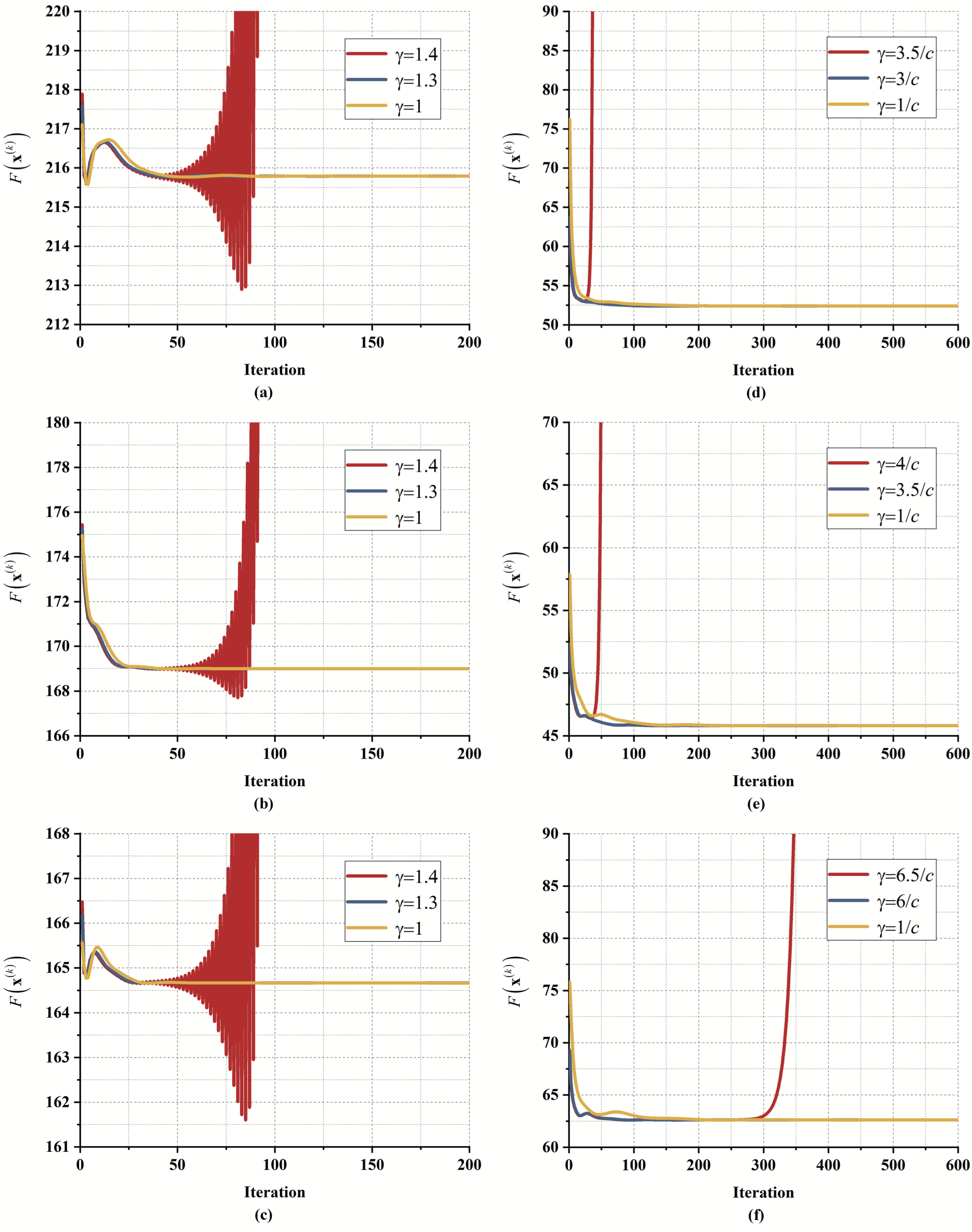}
\caption{Empirical convergences of pFISTA-SENSE and pFISTA-SPIRiT with the hand-tuned optimal and recommended $\gamma$ in terms of function values. (a-c) are the convergences of pFISTA-SENSE for the 8-coil, 12-coil, and 32-coil data. (d-f) are the convergences of pFISTA-SPIRiT for the 8-coil, 12-coil, and 32-coil data. All experiments used the same sampling pattern depicted in Fig. \ref{fig_Dataset}.}%
\label{gamma_gap_cost}
\end{figure}

\subsection{Recommended $\gamma$ and Hand-Tuned Optimal $\gamma$}
One would concern about the gap between the proved sufficient condition and the ground-truth optimal $\gamma$. Here we swept a series of $\gamma$ with a $0.1$ interval for pFISTA-SENSE and a $0.5$ interval for pFISTA-SPIRiT to determine a hand-tuned optimal $\gamma$. We take this hand-tuned optimal $\gamma$ as the ground-truth $\gamma$ to discuss how much is the room lied between the recommended $\gamma$ and the ground-truth $\gamma$. Also, we discuss the influence of the gap on the convergence speed.

\begin{figure}[htb]
\setlength{\abovecaptionskip}{0.cm}
\setlength{\belowcaptionskip}{-0.cm}
\centering
\includegraphics[width=3.5in]{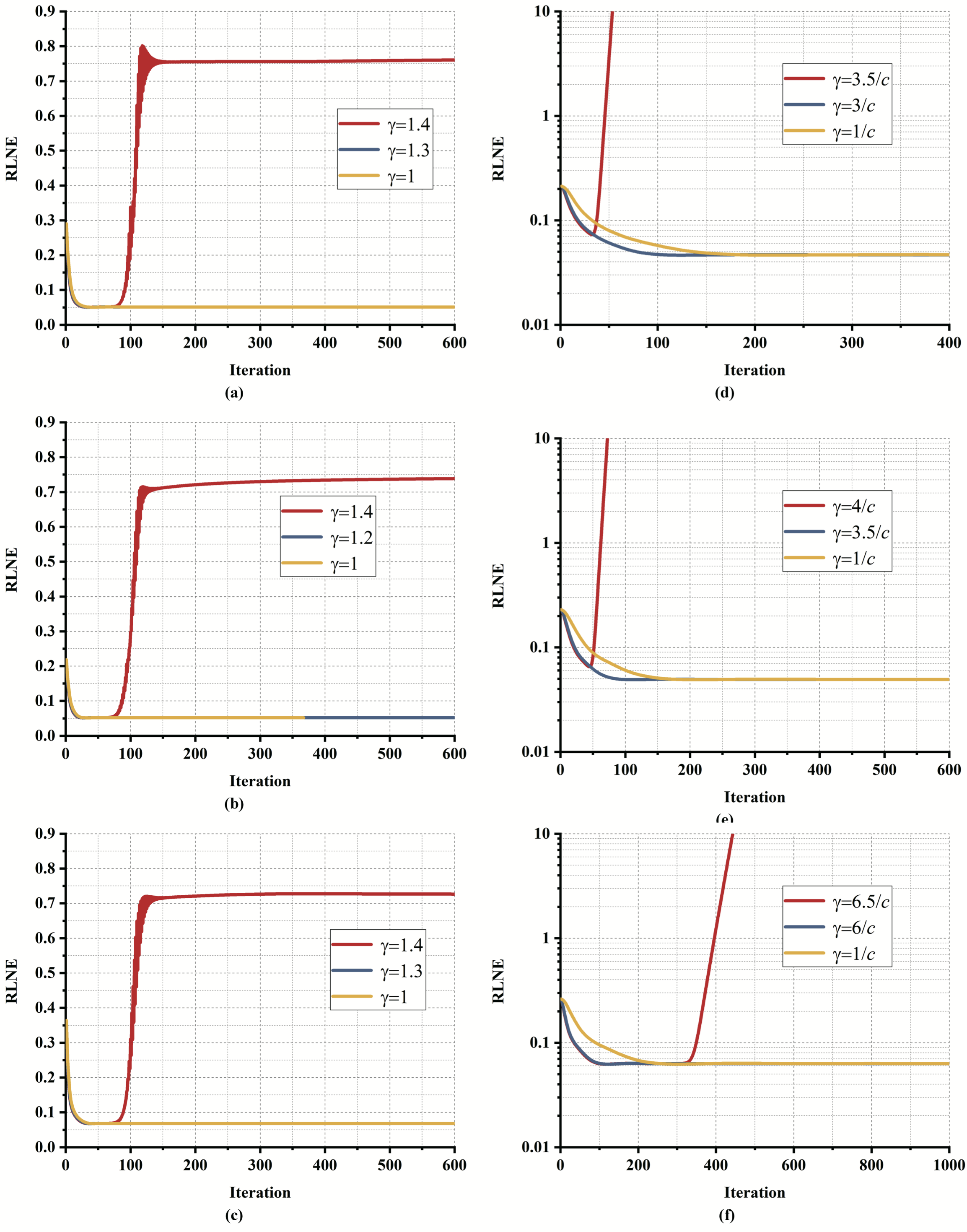}
\caption{Empirical convergences of pFISTA-SENSE and pFISTA-SPIRiT with the hand-tuned optimal and recommended $\gamma$ in terms of RLNE. (a-c) are the convergences of pFISTA-SENSE for the 8-coil, 12-coil, and 32-coil data. (d-f) are the convergences of pFISTA-SPIRiT for the 8-coil, 12-coil, and 32-coil data. All experiments used the same sampling pattern depicted in Fig. \ref{fig_Dataset}.}%
\label{gamma_gap_RLNE}
\end{figure}

In pFISTA-SENSE, the hand-tuned optimal $\gamma=1.3$, which is very close to the recommended $\gamma=1$ (Figs. \ref{gamma_gap_cost} (a-c)). Notably, the two convergence speeds with $\gamma = 1.3$ and $\gamma=1$ are very close. Besides, they used almost the same amount of time to approach the final RLNE level (Figs. \ref{gamma_gap_RLNE} (a-c))). Notably, our recommended $\gamma$ for SENSE is $1$ being independent of the number of coils.

In pFISTA-SPIRiT, as shown in Eq. \eqref{(pFSITA-SPIRiT C-2)}, the recommended $\gamma$ depends on the number of coils. Moreover, the experimental results shown in Figs. \ref{gamma_gap_cost} (d-f) are consistent with the results indicated by Eq. \eqref{(pFSITA-SPIRiT C-2)}. The gap between the recommended $\gamma$ and the hand-tuned optimal $\gamma$ increases as the increase of the number of the coils. For example, the recommended $\gamma$ is three-times smaller than the hand-tuned optimal gamma for the 8-coil image reconstruction. Nevertheless, the recommended $\gamma$ is about six times smaller than the hand-crafted $\gamma$ when the number of coils reaches $32$. Importantly, even there exists a six-fold gap, no distinct difference between their convergence speeds. The hand-tuned optimal $\gamma$ just spends slightly less iterations than the recommended $\gamma$ (Figs. \ref{gamma_gap_RLNE} (d-f)).

\begin{figure}[!htb]
\setlength{\abovecaptionskip}{0.cm}
\setlength{\belowcaptionskip}{-0.cm}
\centering
\includegraphics[width=3.5in]{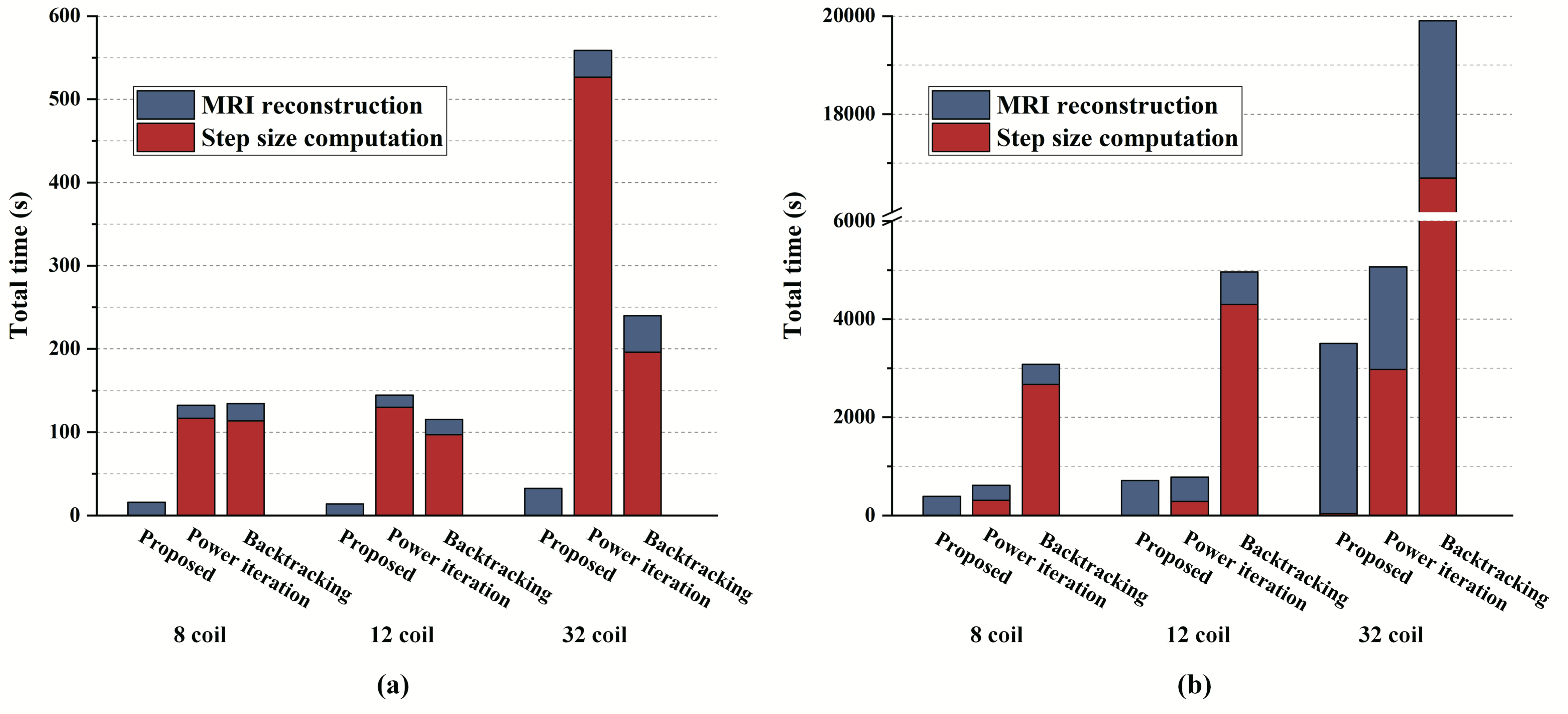}
\caption{The MRI reconstruction and step size computation runtime of power iteration, backtracking, and the recommended step size for pFISTA-SENSE (a) and pFISTA-SPIRiT (b). Note: The "MRI reconstruction" denotes the time spent on iterative reconstruction of an MRI image. The "Step size computation" is the time spent on computing $\gamma$. All experiments used the same sampling pattern depicted in Fig. \ref{fig_Dataset}.}
\label{BT_PI_pFISTA}
\end{figure}

\subsection{Compare with Other Methods to Determine $\gamma$}
One of the essential components of this work is to compute the step size, $\gamma$. It is helpful to consider backtracking and power iteration when the Lipschitz constant is unknown or hard to compute directly. Therefore, we conduct experiments to evaluate the performance of the three methods of computing $\gamma$. Among these three methods, backtracking needs to seek $\gamma$ at every iteration, whereas power iteration and the proposed method (pFISTA-SPIRiT) only need to compute $\gamma$ once before reconstruction. Notably, the recommended sufficient condition of pFISTA for SENSE is 1, which means there is no need to compute $\gamma$.

For SENSE reconstruction (Fig. \ref{BT_PI_pFISTA} (a)), pFISTA allows us to save the time used to compute $\gamma$ by backtracking and power iteration. At least $100$ seconds are save for the tested 8-coil images, and the time will be longer as the number of the coils increases. When considering the total runtime of the program (Step size computation time + MRI reconstruction time), we can see that the recommended $\gamma$ enables pFISTA to have more than five times faster total runtime than backtracking and power iteration. For SPIRiT reconstructions (Fig. \ref{BT_PI_pFISTA} (b)), the computational time of $\gamma$ of pFISTA is also much shorter than that of backtracking and power iteration. As pFISTA spends a relatively long time for reconstruction than power iteration, pFISTA allows only slightly time acceleration over power iteration in terms of the total runtime. However, pFISTA permits more than $5$ times faster runtime than backtracking.

\begin{figure}[htb]
\setlength{\abovecaptionskip}{0.cm}
\setlength{\belowcaptionskip}{-0.cm}
\centering
\includegraphics[width=3.4in]{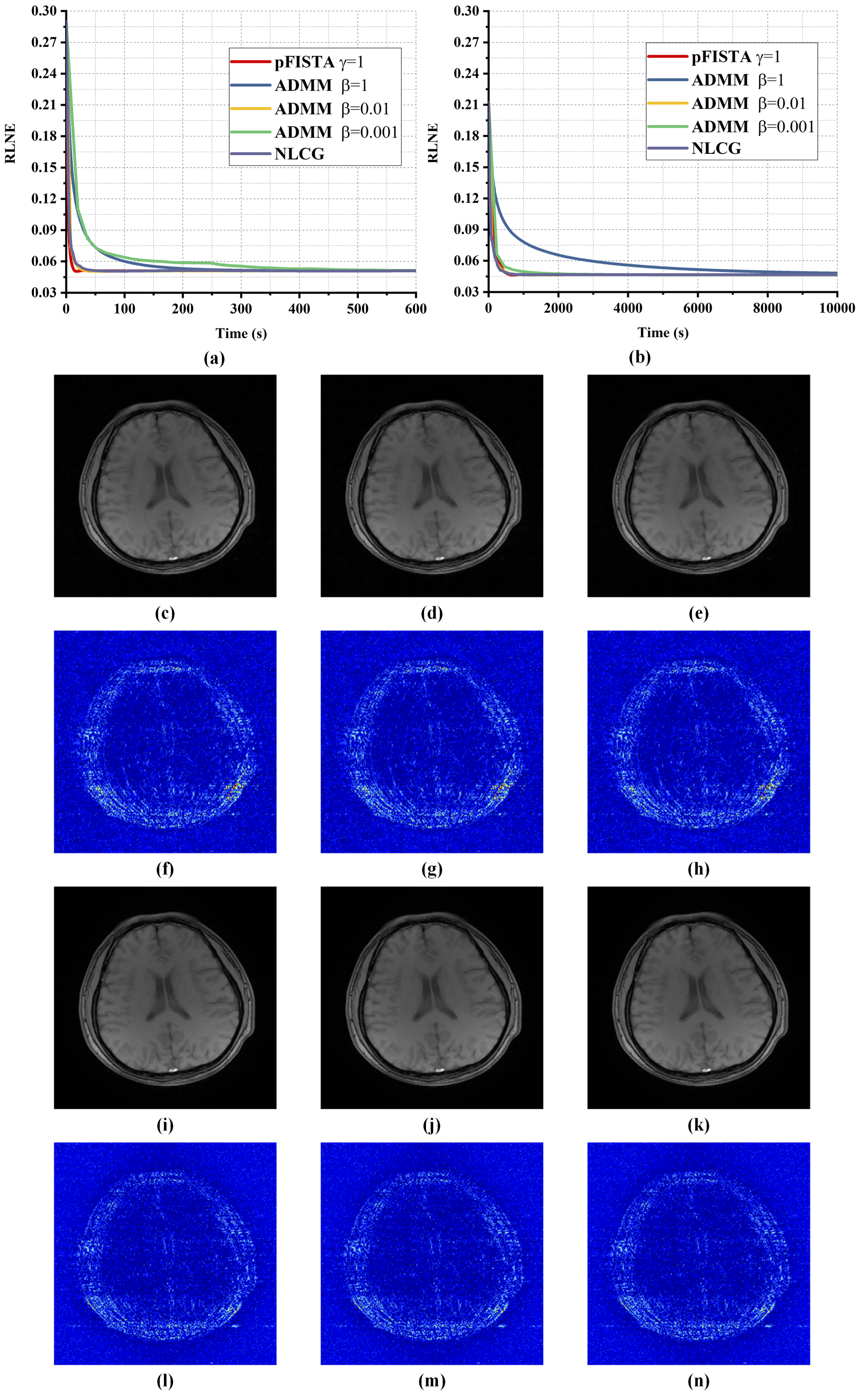}
\caption{Reconstruction results of ADMM, NLCG and pFISTA. (a) (or (b)) are convergence of three algorithms under SENSE-based (or SPIRiT-based) reconstructions in terms of RLNE. (c-e) (or (i-k)) are reconstruction images by ADMM ($\beta=0.01$), NLCG and pFISTA under SENSE-based (or SPIRiT-based) reconstruction. (f-h) and (l-n) are the reconstruction error distribution ($10 \times$) corresponding to reconstructed image above them. Note: 8-coil image in Fig. \ref{fig_Dataset} (a) and the Cartesian sampling pattern shown in Fig. \ref{fig_Dataset} (d) are adopted in all experiments.}
\label{fig_Compared_Method}
\end{figure}
\subsection{Comparison with Other Popular Algorithms}
Here we compare some popular approaches for solving analysis models, including ADMM and NLCG. The results show that ADMM with step size $\beta = 0.01$, NLCG and pFISTA spend almost the same amount of time to reach the final RLNE level (Fig. \ref{fig_Compared_Method}). Moreover, their reconstruction images are similar yielding comparable reconstruction errors. However, compared with pFISTA, ADMM consumes more memory during the reconstruction. Despite NLCG has similar convergence speed in terms of RLNE as pFISTA, the NLCG has more than one parameter to set (for searching step size).
It is worthy to point out that, as shown in Fig. \ref{fig_Compared_Method}, the reconstruction of the 8-coil $T_{1}$-weighted brain image, the convergence of ADMM (in terms of RLNE) is sensitive to the parameter $\beta$ selection, relatively larger or smaller $\beta$ would result in noticeable discrepancy (Figs. \ref{fig_Compared_Method} (a-b)). Furthermore, $\beta=0.01$ yields the fastest convergence speed of ADMM. In summary, pFISTA still holds advantages over these algorithms, such as costing fewer memories, having only one parameter to tune. 

\begin{figure}[htb]
\setlength{\abovecaptionskip}{0.cm}
\setlength{\belowcaptionskip}{-0.cm}
\centering
\includegraphics[width=3.4in]{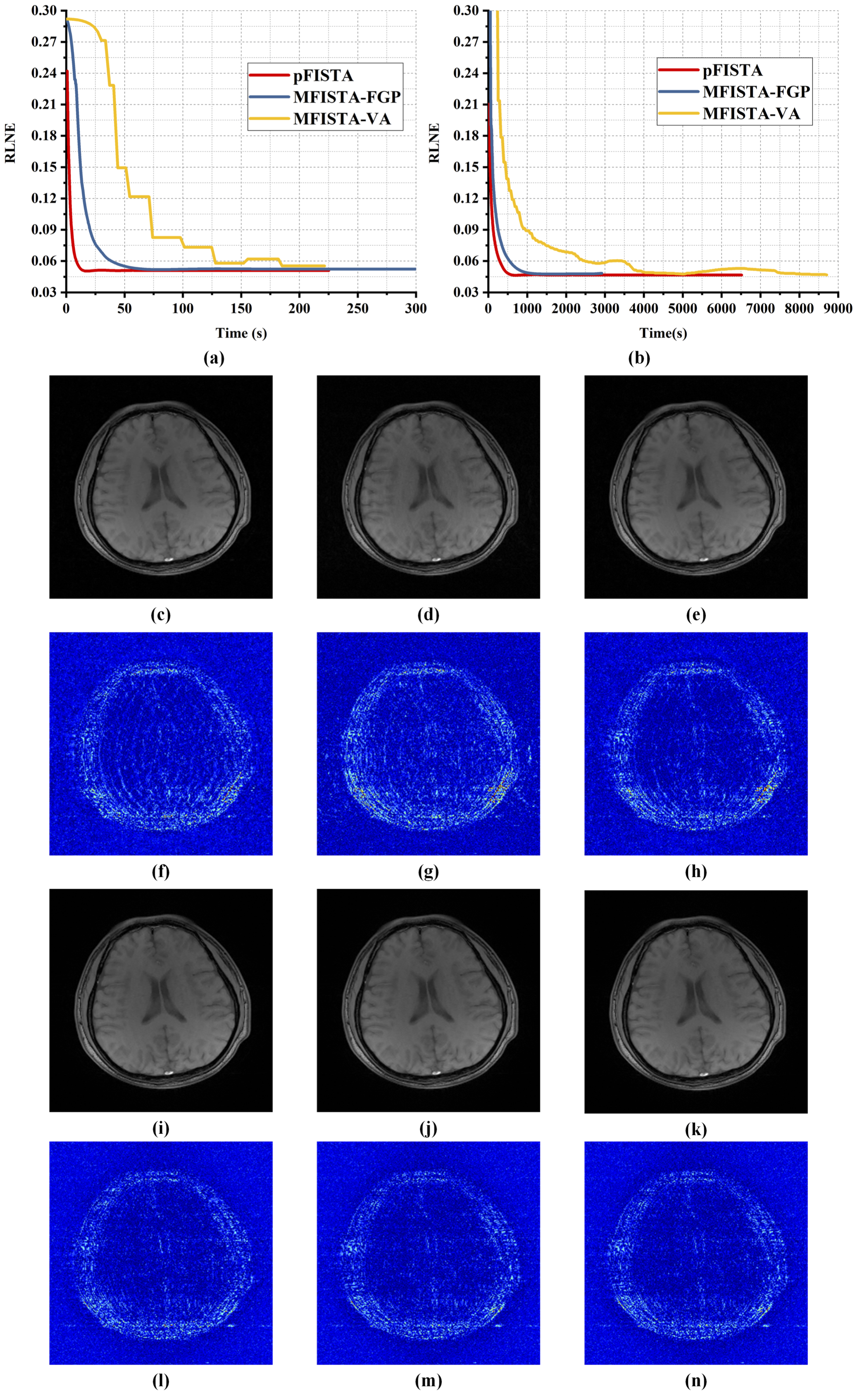}
\caption{Reconstruction results of variants of FISTA and pFISTA. (a) (or (b)) are empirical convergence of three algorithms under SENSE-based (or SPIRiT-based) reconstructions in terms of RLNE. (c-e) (or (i-k)) are reconstruction images by MFISTA-FGP, MFISTA-VA and pFISTA under SENSE-based (or SPIRiT-based) reconstruction. (f-h) and (l-n) are the reconstruction error distribution ($10 \times$) corresponding to reconstructed image above them. Note: 8-coil image in Fig. \ref{fig_Dataset} (a) and the Cartesian sampling pattern shown in Fig. \ref{fig_Dataset} (d) are adopted in all experiments.}%
\label{compare_other_FISTA}
\end{figure}
\subsection{Comparison with Other FISTA Algorithms}
Here, we carried out experiments using other variants of FISTA for solving the analysis model, including MFISTA-FGP \cite{2009_TIP_FISTA_FGP} and MFISTA-VA \cite{2018_TCI_MFISTA_VA}, and the proposed approach. We modified the system matrix of MFISTA-FGP and MFISTA-VA to make it support SENSE and SPIRiT reconstructions.

The results shown in Fig. \ref{compare_other_FISTA} indicate that for both SENSE and SPIRiT reconstructions, pFISTA enables the shortest reconstruction time compared to MFISTA-FGP and MFISTA-VA in terms of RLNE. MFISTA-VA costs the longest time to approach the final RLNE level (Figs. \ref{compare_other_FISTA} (a-b)).

Furthermore, from the RLNE curves and the reconstructed images (Fig. \ref{compare_other_FISTA}), we can see that pFISTA could offer slightly lower RLNEs than MFISTA-FGP and MFISTA-VA, which may enable better reconstruction images. For the SENSE reconstruction shown in Figs. \ref{compare_other_FISTA} (c-h), the MFISTA-FGP error image exhibits noticeable undersampling artifacts inside the skull while MFISTA-VA and pFISTA provide good artifacts suppression. The difference between MFISTA-VA and pFISTA is not so big, but we can still know that pFISTA produces lower reconstruction error. For SPIRiT reconstructions, the three methods offer very close RLNEs, and the reconstructed images are almost the same.

\subsection{Discussion on Other Tight Frames}
In this section, we conduct experiments using pFISTA-SENSE and pFISTA-SPIRiT with SIDWT and four other tight frames, contourlet \cite{2005_TIP_Contourlet, 2010_IPSE_Contourlet_MRI}, shearlet \cite{2008_ACHA_Shearlet}, patch-based directional wavelets (PBDW) \cite{2012_MRI_PBDW}, and PBDW in SIDWT domain (PBDWS) \cite{2013_MRI_PBDWS}. The experimental results demonstrate that the selection of different tight frames will not affect the convergence conditions and for different tight frames, $\gamma = 1/c$ still enables the fastest convergence speed in both SENSE-based and SPIRiT-based recosntruction (in SENSE-based reconstruction $c=1$). Besides, adaptive tight frames, such as PBDW and PBDWS, offer better reconstruction than the pre-defined tight frames like SIDWT, contourlet, and shearlet.

%----------------------------------------------------------------------
%----------------------------- Conclusion -----------------------------
%----------------------------------------------------------------------
\section{Conclusion}\label{Section:conclusion}
As a simple and fast algorithm to solve the sparse reconstruction model, pFISTA has been successfully extended to solve parallel MR imaging problems, but its convergence criterion needs to be proved to help users quickly and conveniently determine the only parameter - step size. Besides, the convergence analysis of single-coil pFISTA cannot be applied to the multi-coil pFISTA. In this work, we provide the guaranteed convergence analysis for parallel imaging version pFISTA to solve spare reconstruction models. More explicitly, along with the sufficient condition, we offer recommended step sizes for both SENSE and SPIRiT. Experimental results evince the validity and effectiveness of the convergence criterion. Further, the recommended step sizes provide more than five times faster reconstruction time in most tested experiments when comparing with the backtracking and power iteration. This work is expected to help users quickly choose the step size to obtain faithful results and fast convergence speed and to promote the application of sparse reconstruction in parallel MRI.

%----------------------------------------------------------------------
%--------------------------- Acknowledgments --------------------------
%----------------------------------------------------------------------
\section*{Acknowledgments}
The authors are grateful to the reviewers and editors for their constructive comments which help improve the writing, convergence analysis, and comparisons with related algorithms. The authors appreciate the help of Yunsong Liu for revising the manuscript. Xiaobo Qu is grateful to Prof. Chun Yuan for hosting his visit to the University of Washington.

%----------------------------------------------------------------------
%----------------------------- References -----------------------------
%----------------------------------------------------------------------
\ifCLASSOPTIONcaptionsoff
  \newpage
\fi

% \newpage
\bibliographystyle{IEEEtran}
\bibliography{IEEEabrv,Mylib}

\begin{thebibliography}{10}
\providecommand{\url}[1]{#1}
\csname url@samestyle\endcsname
\providecommand{\newblock}{\relax}
\providecommand{\bibinfo}[2]{#2}
\providecommand{\BIBentrySTDinterwordspacing}{\spaceskip=0pt\relax}
\providecommand{\BIBentryALTinterwordstretchfactor}{4}
\providecommand{\BIBentryALTinterwordspacing}{\spaceskip=\fontdimen2\font plus
\BIBentryALTinterwordstretchfactor\fontdimen3\font minus
  \fontdimen4\font\relax}
\providecommand{\BIBforeignlanguage}[2]{{%
\expandafter\ifx\csname l@#1\endcsname\relax
\typeout{** WARNING: IEEEtran.bst: No hyphenation pattern has been}%
\typeout{** loaded for the language `#1'. Using the pattern for}%
\typeout{** the default language instead.}%
\else
\language=\csname l@#1\endcsname
\fi
#2}}
\providecommand{\BIBdecl}{\relax}
\BIBdecl

\bibitem{2006_TIT_CS}
D.~L. Donoho, ``Compressed sensing,'' \emph{IEEE Transactions on Information
  Theory}, vol.~52, no.~4, pp. 1289--1306, 2006.

\bibitem{2006_TIT_Tao}
T.~T. E.~J.~Cand{\`e}s, J.~Romberg, ``Robust uncertainty principles: {E}xact
  signal reconstruction from highly incomplete frequency information,''
  \emph{IEEE Transactions on Information Theory}, vol.~52, no.~2, pp. 489--509,
  2006.

\bibitem{2007_MRM_Sparse_MRI}
J.~M.~P. M.~Lustig, D.~Donoho, ``Sparse {MRI}: {T}he application of compressed
  sensing for rapid {MR} imaging,'' \emph{Magnetic Resonance in Medicine},
  vol.~58, no.~6, pp. 1182--1195, 2007.

\bibitem{2008_Qu_contourlet}
X.~{Qu}, {Di Guo}, {Zhong Chen}, and {Congbo Cai}, ``Compressed sensing {MRI}
  based on nonsubsampled contourlet transform,'' in \emph{2008 IEEE
  International Symposium on IT in Medicine and Education}, 2008, pp. 693--696.

\bibitem{2010_IPSE_Contourlet_MRI}
X.~Qu, W.~Zhang, D.~Guo, C.~Cai, S.~Cai, and Z.~Chen, ``Iterative thresholding
  compressed sensing {MRI} based on contourlet transform, inverse problems in
  science and engineering,'' \emph{Inverse Problems in Science and
  Engineering}, vol.~18, no.~6, pp. 737--758, 2010.

\bibitem{2011_TMI_Redundant_wavelet_recon}
M.~Guerquin-Kern, M.~Haberlin, K.~P. Pruessmann, and M.~Unser, ``A fast
  wavelet-based reconstruction method for magnetic resonance imaging,''
  \emph{IEEE Transactions on Medical Imaging}, vol.~30, no.~9, pp. 1649--1660,
  2011.

\bibitem{2011_SIDWT_Ying}
C.~A. Baker, K.~King, D.~Liang, and L.~Ying, ``Translational-invariant
  dictionaries for compressed sensing in magnetic resonance imaging,'' in
  \emph{2011 IEEE International Symposium on Biomedical Imaging: From Nano to
  Macro}, Conference Proceedings, pp. 1602--1605.

\bibitem{2011_TMI_Redundant_Sai}
S.~Ravishankar and Y.~Bresler, ``{MR} image reconstruction from highly
  undersampled k-space data by dictionary learning,'' \emph{IEEE Transactions
  on Medical Imaging}, vol.~30, no.~5, pp. 1028--1041, 2011.

\bibitem{2011_MIA_Orthogonal_1}
J.~Huang, S.~Zhang, and D.~Metaxas, ``Efficient {MR} image reconstruction for
  compressed {MR} imaging,'' \emph{Medical Image Analysis}, vol.~15, no.~5, pp.
  670--679, 2011.

\bibitem{2012_MRI_PBDW}
X.~Qu, D.~Guo, B.~Ning, Y.~Hou, Y.~Lin, S.~Cai, and Z.~Chen, ``Undersampled
  {MRI} reconstruction with patch-based directional wavelets,'' \emph{Magnetic
  Resonance Imaging}, vol.~30, no.~7, pp. 964--977, 2012.

\bibitem{2014_MIA_PANO}
X.~Qu, Y.~Hou, F.~Lam, D.~Guo, J.~Zhong, and Z.~Chen, ``Magnetic resonance
  image reconstruction from undersampled measurements using a patch-based
  nonlocal operator,'' \emph{Medical Image Analysis}, vol.~18, no.~6, pp.
  843--856, 2014.

\bibitem{2008_Orthogonal_2}
S.~Ma, W.~Yin, Y.~Zhang, and A.~Chakraborty, ``An efficient algorithm for
  compressed {MR} imaging using total variation and wavelets,'' in \emph{2008
  IEEE Conference on Computer Vision and Pattern Recognition}.\hskip 1em plus
  0.5em minus 0.4em\relax IEEE, Conference Proceedings, pp. 1--8.

\bibitem{2007_PSF_liang}
Z.~{Liang}, ``Spatiotemporal imaging with partially separable functions,'' in
  \emph{2007 4th IEEE International Symposium on Biomedical Imaging: From Nano
  to Macro}, 2007, pp. 988--991.

\bibitem{2010_PJO_Low_rank}
K.-C. Toh and S.~Yun, ``An accelerated proximal gradient algorithm for nuclear
  norm regularized linear least squares problems,'' \emph{Pacific Journal of
  Optimization}, vol.~6, no. 615-640, p.~15, 2010.

\bibitem{2015_MRM_Low_rank}
T.~Zhang, J.~M. Pauly, and I.~R. Levesque, ``Accelerating parameter mapping
  with a locally low rank constraint,'' \emph{Magnetic Resonance in Medicine},
  vol.~73, no.~2, pp. 655--661, 2015.

\bibitem{2016_liang_LR}
J.~{He}, Q.~{Liu}, A.~G. {Christodoulou}, C.~{Ma}, F.~{Lam}, and Z.~{Liang},
  ``Accelerated high-dimensional {MR} imaging with sparse sampling using
  low-rank tensors,'' \emph{IEEE Transactions on Medical Imaging}, vol.~35,
  no.~9, pp. 2119--2129, 2016.

\bibitem{2020_xinlin}
X.~Zhang, D.~Guo, Y.~Huang, Y.~Chen, L.~Wang, F.~Huang, Q.~Xu, and X.~Qu,
  ``Image reconstruction with low-rankness and self-consistency of k-space data
  in parallel {MRI},'' \emph{Medical Image Analysis}, vol.~63, p. 101687, 2020.

\bibitem{2011_TMI_k-t_SLR}
S.~G. {Lingala}, Y.~{Hu}, E.~{DiBella}, and M.~{Jacob}, ``Accelerated dynamic
  {MRI} exploiting sparsity and low-rank structure: k-t {SLR},'' \emph{IEEE
  Transactions on Medical Imaging}, vol.~30, no.~5, pp. 1042--1054, 2011.

\bibitem{2015_MRM_L+S}
R.~Otazo, E.~Cand{\`e}s, and D.~K. Sodickson, ``Low-rank plus sparse matrix
  decomposition for accelerated dynamic {MRI} with separation of background and
  dynamic components,'' \emph{Magnetic Resonance in Medicine}, vol.~73, no.~3,
  pp. 1125--1136, 2015.

\bibitem{2018_MRM_L+S}
M.~V. Zibetti, A.~Sharafi, R.~Otazo, and R.~R. Regatte, ``Accelerating
  3{D}-{T}$_{1 \rho}$ mapping of cartilage using compressed sensing with
  different sparse and low rank models,'' \emph{Magnetic Resonance in
  Medicine}, vol.~80, no.~4, pp. 1475--1491, 2018.

\bibitem{2018_TCI_L+S_FISTA}
C.~Y. Lin and J.~A. Fessler, ``Efficient dynamic parallel {MRI} reconstruction
  for the low-rank plus sparse model,'' \emph{IEEE Transactions on
  Computational Imaging}, vol.~5, no.~1, pp. 17--26, 2018.

\bibitem{2016_TBME_FDLCP}
Z.~Zhan, J.~Cai, D.~Guo, Y.~Liu, Z.~Chen, and X.~Qu, ``Fast multiclass
  dictionaries learning with geometrical directions in {MRI} reconstruction,''
  \emph{IEEE Transactions on Biomedical Engineering}, vol.~63, no.~9, pp.
  1850--1861, 2016.

\bibitem{2016_MIA_GBRWT}
Z.~Lai, X.~Qu, Y.~Liu, D.~Guo, J.~Ye, Z.~Zhan, and Z.~Chen, ``Image
  reconstruction of compressed sensing {MRI} using graph-based redundant
  wavelet transform,'' \emph{Medical Image Analysis}, vol.~27, pp. 93--104,
  2016.

\bibitem{2014_tight_frame}
M.~Vetterli, J.~Kovačević, and V.~K. Goyal, \emph{Foundations of {S}ignal
  {P}rocessing}.\hskip 1em plus 0.5em minus 0.4em\relax Cambridge University
  Press, 2014.

\bibitem{2016_TMI_pFISTA}
Y.~Liu, Z.~Zhan, J.-F. Cai, D.~Guo, Z.~Chen, and X.~Qu, ``Projected iterative
  soft-thresholding algorithm for tight frames in compressed sensing magnetic
  resonance imaging,'' \emph{IEEE Transactions on Medical Imaging}, vol.~35,
  no.~9, pp. 2130--2140, 2016.

\bibitem{2008_CRM_CS}
E.~J. Cand{\`e}s, ``The restricted isometry property and its implications for
  compressed sensing,'' \emph{Comptes Rendus Mathematique}, vol. 346, no. 9-10,
  pp. 589--592, 2008.

\bibitem{2008_TIT_redundant_CS}
H.~Rauhut, K.~Schnass, and P.~Vandergheynst, ``Compressed sensing and redundant
  dictionaries,'' \emph{IEEE Transactions on Information Theory}, vol.~54,
  no.~5, pp. 2210--2219, 2008.

\bibitem{2011_ACHA_redundant_CS}
E.~J. Cand{\`e}s, Y.~C. Eldar, D.~Needell, and P.~Randall, ``Compressed sensing
  with coherent and redundant dictionaries,'' \emph{Applied and Computational
  Harmonic Analysis}, vol.~31, no.~1, pp. 59--73, 2011.

\bibitem{2013_ACHA_analysis_model}
S.~Nam, M.~Davies, M.~Elad, and R.~Gribonval, ``The cosparse analysis model and
  algorithms,'' \emph{Applied and Computational Harmonic Analysis}, vol.~34,
  no.~1, pp. 30--56, 2013.

\bibitem{2015_PlosOne_Balance}
Y.~Liu, J.-F. Cai, Z.~Zhan, D.~Guo, J.~Ye, Z.~Chen, and X.~Qu, ``Balanced
  sparse model for tight frames in compressed sensing magnetic resonance
  imaging,'' \emph{PloS One}, vol.~10, no.~4, p. e0119584, 2015.

\bibitem{2020_jeffery}
J.~A. {Fessler}, ``Optimization methods for magnetic resonance image
  reconstruction: Key models and optimization algorithms,'' \emph{IEEE Signal
  Processing Magazine}, vol.~37, no.~1, pp. 33--40, 2020.

\bibitem{2011_ADMM}
S.~Boyd, N.~Parikh, E.~Chu, B.~Peleato, and J.~Eckstein, ``Distributed
  optimization and statistical learning via the alternating direction method of
  multipliers,'' \emph{Foundations and Trends® in Machine Learning}, vol.~3,
  no.~1, pp. 1--122, 2011.

\bibitem{2010_TMI_AL}
S.~Ramani and J.~A. Fessler, ``Parallel {MR} image reconstruction using
  augmented {L}agrangian methods,'' \emph{IEEE Transactions on Medical
  Imaging}, vol.~30, no.~3, pp. 694--706, 2010.

\bibitem{1983_NESTA_1}
Y.~Nesterov, ``A method for unconstrained convex minimization problem with the
  rate of convergence $o(1/k^2)$,'' \emph{Doklady an ussr}, vol. 269, pp.
  543--547, 1983.

\bibitem{2005_NESTA_2}
------, ``Smooth minimization of non-smooth functions,'' \emph{Mathematical
  Programming}, vol. 103, no.~1, pp. 127--152, 2005.

\bibitem{2013_NESTA_3}
------, ``Gradient methods for minimizing composite functions,''
  \emph{Mathematical Programming}, vol. 140, no.~1, pp. 125--161, 2013.

\bibitem{1956_Douglas-Rachford}
J.~Douglas and H.~H. Rachford, ``On the numerical solution of heat conduction
  problems in two and three space variables,'' \emph{Transactions of the
  American Athematical Society}, vol.~82, no.~2, pp. 421--439, 1956.

\bibitem{2007_JSTSP_Douglas-Rachford_Recovery}
P.~L. {Combettes} and J.~{Pesquet}, ``A {D}ouglas {R}achford splitting approach
  to nonsmooth convex variational signal recovery,'' \emph{IEEE Journal of
  Selected Topics in Signal Processing}, vol.~1, no.~4, pp. 564--574, 2007.

\bibitem{2004_ISTA}
I.~Daubechies, M.~Defrise, and C.~De~Mol, ``An iterative thresholding algorithm
  for linear inverse problems with a sparsity constraint,''
  \emph{Communications on Pure Applied Mathematics}, vol.~57, no.~11, pp.
  1413--1457, 2004.

\bibitem{2009_SIAM_FISTA}
A.~Beck and M.~Teboulle, ``A fast iterative shrinkage-thresholding algorithm
  for linear inverse problems,'' \emph{SIAM Journal on Imaging Sciences},
  vol.~2, no.~1, pp. 183--202, 2009.

\bibitem{2011_TP_variants_FISTA_theory}
M.~Yamagishi and I.~Yamada, ``Over-relaxation of the fast iterative
  shrinkage-thresholding algorithm with variable stepsize,'' \emph{Inverse
  Problems}, vol.~27, no.~10, p. 105008, 2011.

\bibitem{2018_JOTA_variants_FISTA_theory}
A.~B. Taylor, J.~M. Hendrickx, and F.~Glineur, ``Exact worst-case convergence
  rates of the proximal gradient method for composite convex minimization,''
  \emph{Journal of Optimization Theory and Applications}, vol. 178, no.~2, pp.
  455--476, 2018.

\bibitem{2009_TIP_FISTA_FGP}
A.~Beck and M.~Teboulle, ``Fast gradient-based algorithms for constrained total
  variation image denoising and deblurring problems,'' \emph{IEEE Transactions
  on Image Processing}, vol.~18, no.~11, pp. 2419--2434, 2009.

\bibitem{2018_TCI_MFISTA_VA}
M.~V.~W. Zibetti, E.~S. Helou, R.~R. Regatte, and G.~T. Herman, ``Monotone
  fista with variable acceleration for compressed sensing magnetic resonance
  imaging,'' \emph{IEEE Transactions on Computational Imaging}, vol.~5, no.~1,
  pp. 109--119, 2018.

\bibitem{2017_MRM_bFISTA}
S.~T. Ting, R.~Ahmad, N.~Jin, J.~Craft, J.~Serafim~da Silveira, H.~Xue, and
  O.~P. Simonetti, ``Fast implementation for compressive recovery of highly
  accelerated cardiac cine {MRI} using the balanced sparse model,''
  \emph{Magnetic Resonance in Medicine}, vol.~77, no.~4, pp. 1505--1515, 2017.

\bibitem{2020_Ahmad}
R.~{Ahmad}, C.~A. {Bouman}, G.~T. {Buzzard}, S.~{Chan}, S.~{Liu}, E.~T.
  {Reehorst}, and P.~{Schniter}, ``Plug-and-play methods for magnetic resonance
  imaging: Using denoisers for image recovery,'' \emph{IEEE Signal Processing
  Magazine}, vol.~37, no.~1, pp. 105--116, 2020.

\bibitem{1999_MRM_SENSE}
K.~P. Pruessmann, M.~Weiger, M.~B. Scheidegger, and P.~Boesiger, ``{SENSE}:
  {S}ensitivity encoding for fast {MRI},'' \emph{Magnetic Resonance in
  Medicine}, vol.~42, pp. 952--962, 1999.

\bibitem{2010_MRM_SPIRiT}
M.~Lustig and J.~M. Pauly, ``{SPIR}i{T}: Iterative self-consistent parallel
  imaging reconstruction from arbitrary k-space,'' \emph{Magnetic Resonance in
  Medicine}, vol.~64, no.~2, pp. 457--71, 2010.

\bibitem{2000_Matrix_Analysis}
C.~D. Meyer, \emph{Matrix {A}nalysis and {A}pplied {L}inear {A}lgebra}.\hskip
  1em plus 0.5em minus 0.4em\relax Siam, 2000, vol.~71.

\bibitem{MFISTA_codes}
\BIBentryALTinterwordspacing
M.~Zibetti. (2019) Matlab codes of {MFISTA-FGP} and {MFISTA-VA}. [Online].
  Available:
  \url{https://www.cai2r.net/resources/software/cs-mri-mfista-va-matlab-code}
\BIBentrySTDinterwordspacing

\bibitem{2007_JSENSE}
L.~Ying and J.~Sheng, ``Joint image reconstruction and sensitivity estimation
  in {SENSE} ({JSENSE}),'' \emph{Magnetic Resonance in Medicine}, vol.~57,
  no.~6, pp. 1196--1202, 2007.

\bibitem{1995_SIDWT_Denosing}
R.~R. Coifman and D.~L. Donoho, \emph{Translation-invariant de-noising}.\hskip
  1em plus 0.5em minus 0.4em\relax Springer, 1995, pp. 125--150.

\bibitem{2014_MRI_SIDWT}
M.~H. Kayvanrad, A.~J. McLeod, J.~S. Baxter, C.~A. McKenzie, and T.~M. Peters,
  ``Stationary wavelet transform for under-sampled {MRI} reconstruction,''
  \emph{Magnetic Resonance Imaging}, vol.~32, no.~10, pp. 1353--1364, 2014.

\bibitem{2005_TIP_Contourlet}
M.~N. Do and M.~Vetterli, ``The contourlet transform: an efficient directional
  multiresolution image representation,'' \emph{IEEE Transactions on Image
  Processing}, vol.~14, no.~12, pp. 2091--2106, 2005.

\bibitem{2008_ACHA_Shearlet}
G.~Easley, D.~Labate, and W.-Q. Lim, ``Sparse directional image representations
  using the discrete shearlet transform,'' \emph{Applied and Computational
  Harmonic Analysis}, vol.~25, no.~1, pp. 25--46, 2008.

\bibitem{2013_MRI_PBDWS}
B.~Ning, X.~Qu, D.~Guo, C.~Hu, and Z.~Chen, ``Magnetic resonance image
  reconstruction using trained geometric directions in 2{D} redundant wavelets
  domain and non-convex optimization,'' \emph{Magnetic Resonance Imaging},
  vol.~31, no.~9, pp. 1611--1622, 2013.

\end{thebibliography}

\end{document}

\title{Supplementary Material of ‘‘A Guaranteed Convergence Analysis for the Projected Fast Iterative Soft-Thresholding Algorithm in Parallel MRI’’}

\author{Xinlin~Zhang,
        Hengfa~Lu,
        Di~Guo,
        Lijun~Bao,
        Feng~Huang,
        Qin~Xu,
        Xiaobo~Qu*}

% \markboth{IEEE TRANSACTIONS ON COMPUTATIONAL IMAGING}%
% {Shell \MakeLowercase{\textit{et al.}}: Bare Demo of IEEEtran.cls for IEEE Journals}

\maketitle

%----------------------------------------------------------------------
%----------------------------- Abstract -------------------------------
%----------------------------------------------------------------------
%\begin{abstract}
In this supplement, we provide proof of Eq. (47) in the main text, and reconstructed results of pFISTA-SENSE and pFISTA-SPIRiT under different tight frames.
%\end{abstract}

\IEEEpeerreviewmaketitle
\renewcommand\thefigure{S\arabic{figure}}{}
\renewcommand\theequation{S\arabic{equation}}{}
\renewcommand\thesection{S\arabic{section}}{}

%----------------------------------------------------------------------
%---------------------------- Supplement ------------------------------
%----------------------------------------------------------------------
\section{Proof of Equation (47)}
In this section, we give the proof of Eq. (47) in the main text.

\begin{proof}
When $i=1,\cdots ,z$, where $z$ equals the largest integer no more than $\frac{J}{2}$, the matrix ${{\mathbf{Z}}_{diag,i}}+{{\mathbf{Z}}_{diag,-i}}$ can be converted to diagonal matrices by elementary transforms which do not affect the $\ell_2$ norm of a matrix:
\begin{equation}
	\left[ \begin{matrix}
   {} & {} & {} & {{\mathbf{Z}}_{1,i}} & {} & {}  \\
   {} & {} & {} & {} & \ddots  & {}  \\
   {} & {} & {} & {} & {} & {{\mathbf{Z}}_{J-i+1,J}}  \\
   {{\mathbf{Z}}_{i,1}} & {} & {} & {} & {} & {}  \\
   {} & \ddots  & {} & {} & {} & {}  \\
   {} & {} & {{\mathbf{Z}}_{J,J-i+1}} & {} & {} & {}  \\
\end{matrix} \right]\to \left[ \begin{matrix}
   {{\mathbf{Z}}_{i,1}} & {} & {} & {} & {} & {}  \\
   {} & \ddots  & {} & {} & {} & {}  \\
   {} & {} & {{\mathbf{Z}}_{J,J-i+1}} & {} & {} & {}  \\
   {} & {} & {} & {{\mathbf{Z}}_{1,i}} & {} & {}  \\
   {} & {} & {} & {} & \ddots  & {}  \\
   {} & {} & {} & {} & {} & {{\mathbf{Z}}_{J-i,+1J}}  \\
\end{matrix} \right].
\end{equation}
Then,
\begin{equation}
\begin{aligned}
  & \;\;\;\; {{\left\| {{\mathbf{Z}}_{diag,i}}+{{\mathbf{Z}}_{diag,-i}} \right\|}_{2}}\\
  &=\max \left( \mathbf{Z}_{1,i}^{H}{{\mathbf{Z}}_{1,i}},\cdots ,\mathbf{Z}_{J-i+1,J}^{H}{{\mathbf{Z}}_{J-i+1,J}},\mathbf{Z}_{i,1}^{H}{{\mathbf{Z}}_{i,1}},\cdots ,\mathbf{Z}_{J,J-i+1}^{H}{{\mathbf{Z}}_{J,J-i+1}} \right) \\ 
 & =\max \left( \mathbf{Z}_{1,i}^{H}{{\mathbf{Z}}_{1,i}},\cdots ,\mathbf{Z}_{J-i+1,J}^{H}{{\mathbf{Z}}_{J-i+1,J}},{{\mathbf{Z}}_{1,i}}\mathbf{Z}_{1,i}^{H},\cdots ,{{\mathbf{Z}}_{J-i+1,J}}\mathbf{Z}_{J-i+1,J}^{H} \right).
\end{aligned}
\end{equation}
Since the ${{\mathbf{Z}}_{i,j}}$ is a diagonal matrix, we have ${{\mathbf{Z}}_{i,j}}\mathbf{Z}_{i,j}^{H}=\mathbf{Z}_{i,j}^{H}{{\mathbf{Z}}_{i,j}}$. Therefore,
\begin{equation}\label{(Z_daig,1+Z_diag,-i)}
	\begin{aligned}
  & \;\;\;\; {{\left\| {{\mathbf{Z}}_{diag,i}}+{{\mathbf{Z}}_{diag,-i}} \right\|}_{2}}\\
  &=\max \left( \mathbf{Z}_{1,i}^{H}{{\mathbf{Z}}_{1,i}},\cdots ,\mathbf{Z}_{J-i+1,J}^{H}{{\mathbf{Z}}_{J-i+1,J}},{{\mathbf{Z}}_{1,i}}\mathbf{Z}_{1,i}^{H},\cdots ,{{\mathbf{Z}}_{J-i+1,J}}\mathbf{Z}_{J-i+1,J}^{H} \right) \\ 
 & =\max \left( \mathbf{Z}_{1,i}^{H}{{\mathbf{Z}}_{1,i}},\cdots ,\mathbf{Z}_{J-i+1,J}^{H}{{\mathbf{Z}}_{J-i+1,J}} \right). 
\end{aligned}
\end{equation}
Besides, since the matrix ${{\mathbf{Z}}_{i,j}}$ is a diagonal matrix, the matrix ${{\mathbf{Z}}_{diag,i}}$ can also be easily converted to a diagonal matrix by elementary transforms. Thus, we have
\begin{equation}\label{(Z_diag,i)}
	{{\left\| {{\mathbf{Z}}_{diag,i}} \right\|}_{2}}=\max \left( \mathbf{Z}_{1,i}^{H}{{\mathbf{Z}}_{1,i}},\cdots ,\mathbf{Z}_{J-i+1,J}^{H}{{\mathbf{Z}}_{J-i+1,J}} \right).
\end{equation}
Notice that the right-hand side of Eq. \eqref{(Z_diag,i)} and that of Eq. \eqref{(Z_daig,1+Z_diag,-i)} are the same, thus,
\begin{equation}
	{{\left\| {{\mathbf{Z}}_{diag,i}}+{{\mathbf{Z}}_{diag,-i}} \right\|}_{2}}={{\left\| {{\mathbf{Z}}_{diag,i}} \right\|}_{2}}.
\end{equation}
\end{proof}

\newpage
\section{Discussion on other tight frames}
The tight frame is crucial for sparse MRI reconstruction. In this section, we conduct experiments using pFISTA-SENSE and pFISTA-SPIRiT with SIDWT and four other tight frames, contourlet \cite{2005_TIP_Contourlet, 2010_IPSE_Contourlet_MRI}, shearlet \cite{2008_ACHA_Shearlet}, PBDW \cite{2012_MRI_PBDW}, and PBDWS \cite{2013_MRI_PBDWS}. By exploring the image geometry, contourlet provides a sparse expansion for images that have smooth contours \cite{2005_TIP_Contourlet}, and is applied to preserve smooth edges on MRI reconstruction \cite{2010_IPSE_Contourlet_MRI}. Shearlet combines the ability to capture the geometric features of data with the power of multi-scale methods and therefore provides improvements compared to the contourlet transform \cite{2008_ACHA_Shearlet}. PBDW trains the geometric directions on the pixels of image patches and provides an adaptively sparse representation for image \cite{2012_MRI_PBDW}. PBDWS extends the PBDW into the SIDWT domain to enhance the ability of sparsifying \cite{2013_MRI_PBDWS}. Here, the filters used in contourlet are ladder structure filters, and the decomposition levels are $[5, 4, 4, 3]$, the filters used in shearlet are Meyer filters \cite{2001_Meyer}, and the decomposition level used in shearlet is $2$, and the filters used in PBDW and PBDWS are the Haar wavelets and the decomposition level is $3$.

\begin{figure*}[!htb]
\setlength{\abovecaptionskip}{0.cm}
\setlength{\belowcaptionskip}{0.cm}
\centering
\includegraphics[width=7.2in]{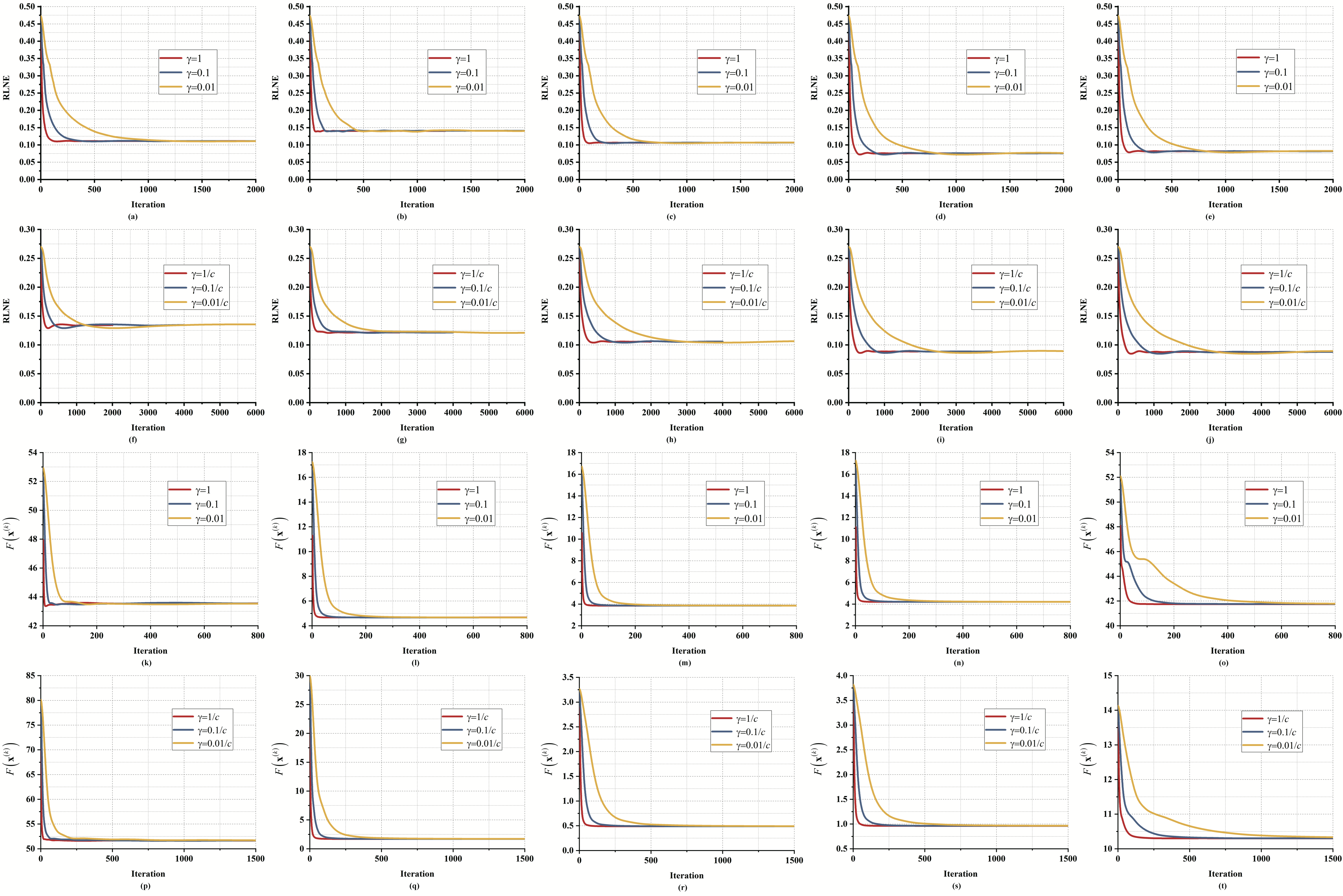}
\caption{Empirical convergence using five tight frames under a sampling rate of $0.25$. (a-e) (or (f-j)) are the RLNEs of pFISTA-SENSE (or pFISTA-SPIRiT) with different $\gamma $ using SIDWT, contourlet, shearlet, PBDW, and PBDWS, respectively; and (k-o) (or (p-t)) are the objective function values of pFISTA-SENSE (or pFISTA-SPIRiT) with different $\gamma $ using SIDWT, contourlet, shearlet, PBDW, and PBDWS, respectively. Note: 8-coil image in Fig. 3 (a) and the Cartesian sampling pattern with sampling rate of $0.25$ are adopted in all experiments.}
\label{fig_psi_convergence}
\end{figure*}

\begin{figure*}[!htb]
\setlength{\abovecaptionskip}{0.cm}
\setlength{\belowcaptionskip}{0.cm}
\centering
\includegraphics[width=7.0in]{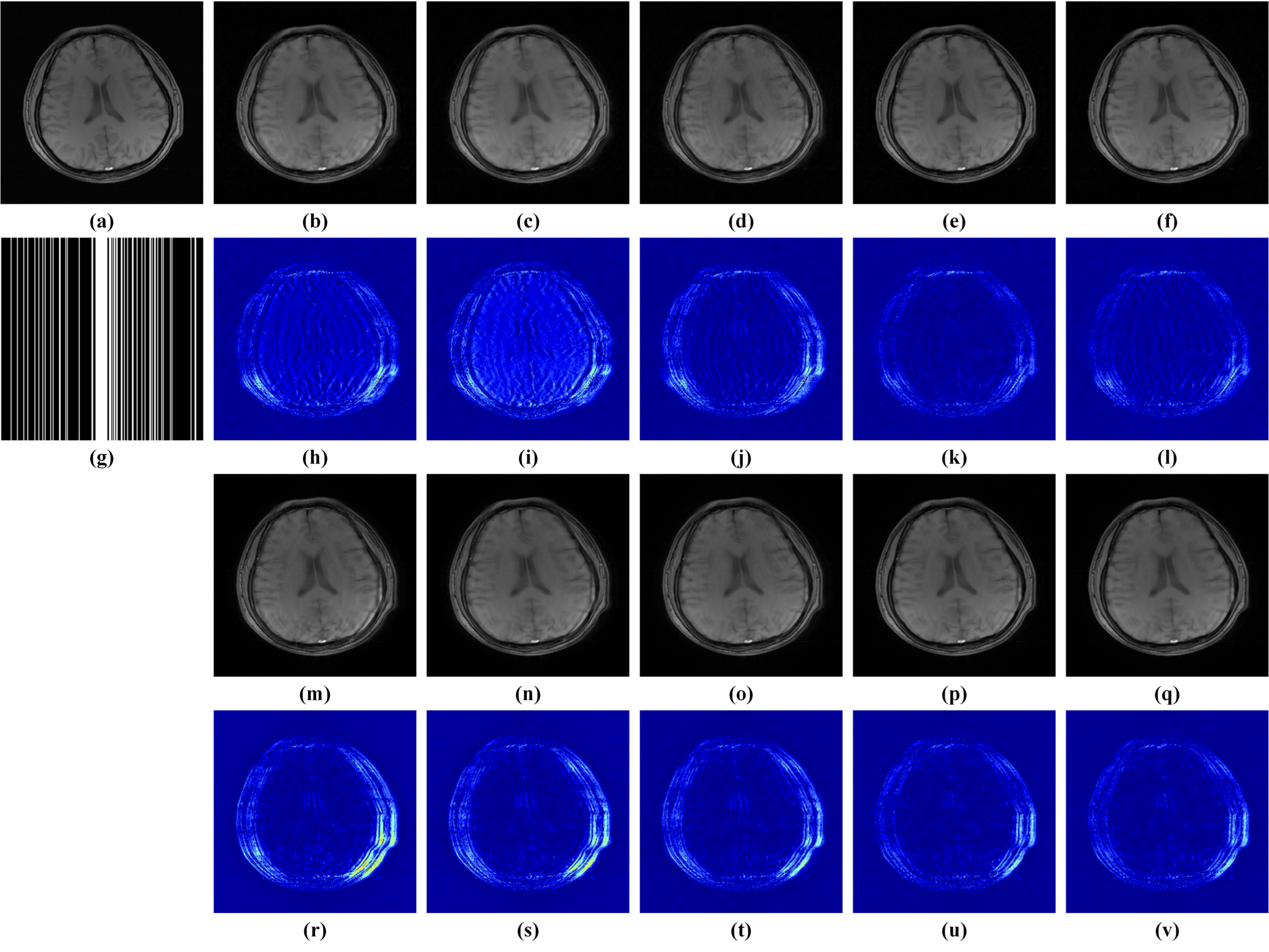}
\caption{Reconstruction results of SIDWT and other four tight frames under a sampling rate of $0.25$. (a) is the fully sampled image; (b-f) (or (m-q)) are images of reconstruction results of pFISTA-SENSE with $\gamma =1$ (or pFISTA-SPIRiT with $\gamma=1/c$) using SIDWT, contourlet, shearlet, PBDW, and PBDWS, respectively; (g) is 1D Cartesian sampling pattern with a sampling rate of $0.25$ and $14$ ACS lines acquired ; (h-l) (or (r-v)) are the reconstruction error distributions ($3\times$) corresponding to the reconstructed images above them.}
\label{fig_psi_images}
\end{figure*}

The experimental results in Fig. \ref{fig_psi_convergence} confirm our argument that the recommended sufficient condition $\gamma = 1/c$ enables the fastest convergence speed in both SENSE-based and SPIRiT reconstructions. Both the RLNE and cost function curves show that $\gamma = 1/c$ allows the fastest convergence speed. This result indicates that the sufficient condition is robust to different tight frames, permitting flexibility in applying the sufficient conditions in practical scenarios. Besides, the reconstructed images in Fig. \ref{fig_psi_images} show that PBDW and PBDWS provide much better reconstructions than the rest of the tight frames in terms of lower reconstruction errors and artifacts suppression. The results are reasonable since PBDW and PBDWS are tight frames trained using a pre-reconstructed image, which will offer a sparser representation of the image. 

\begin{figure}[H]
\setlength{\abovecaptionskip}{0.cm}
\setlength{\belowcaptionskip}{-0.cm}
\centering
\includegraphics[width=6.5in]{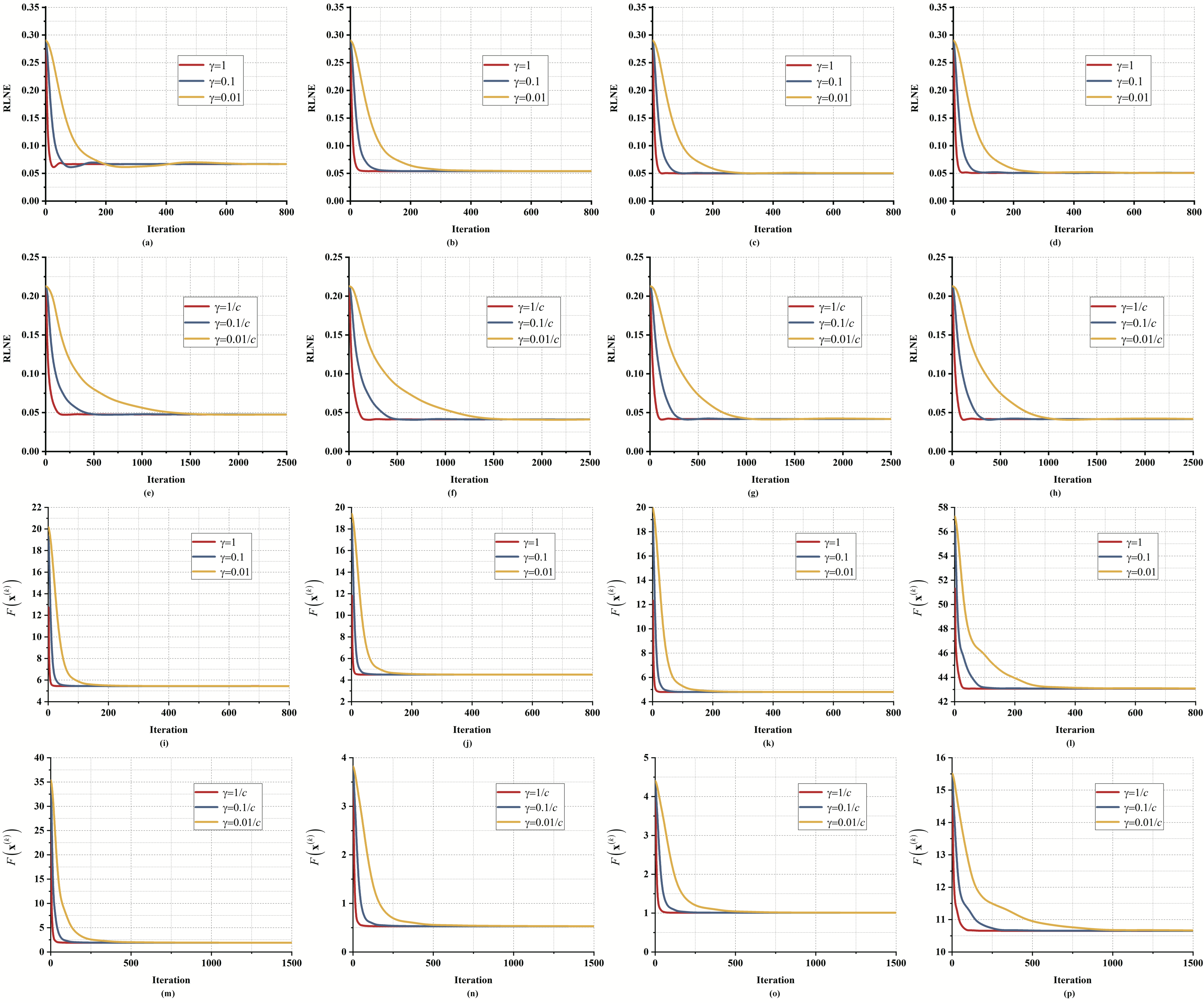}
\caption{Empirical convergence using four other tight frames under a sampling rate of $0.34$. (a-d) (or (e-h)) are the RLNEs of pFISTA-SENSE (or pFISTA-SPIRiT) with different $\gamma$ using contourlet, shearlet, PBDW, and PBDWS, respectively; and (i-l) (or (m-p)) are the function values of pFISTA-SENSE (or pFISTA-SPIRiT) with different $\gamma$ using contourlet, shearlet, PBDW, and PBDWS, respectively. Note: 8-coil image in Fig. 3 (a) in the main text and sampling pattern in Fig. 3 (d) in the main text are adopted in all experiments.}
\label{fig_Convergence34}
\end{figure}

\begin{figure}[H]
\setlength{\abovecaptionskip}{0.cm}
\setlength{\belowcaptionskip}{-0.cm}
\centering
\includegraphics[width=5.0in]{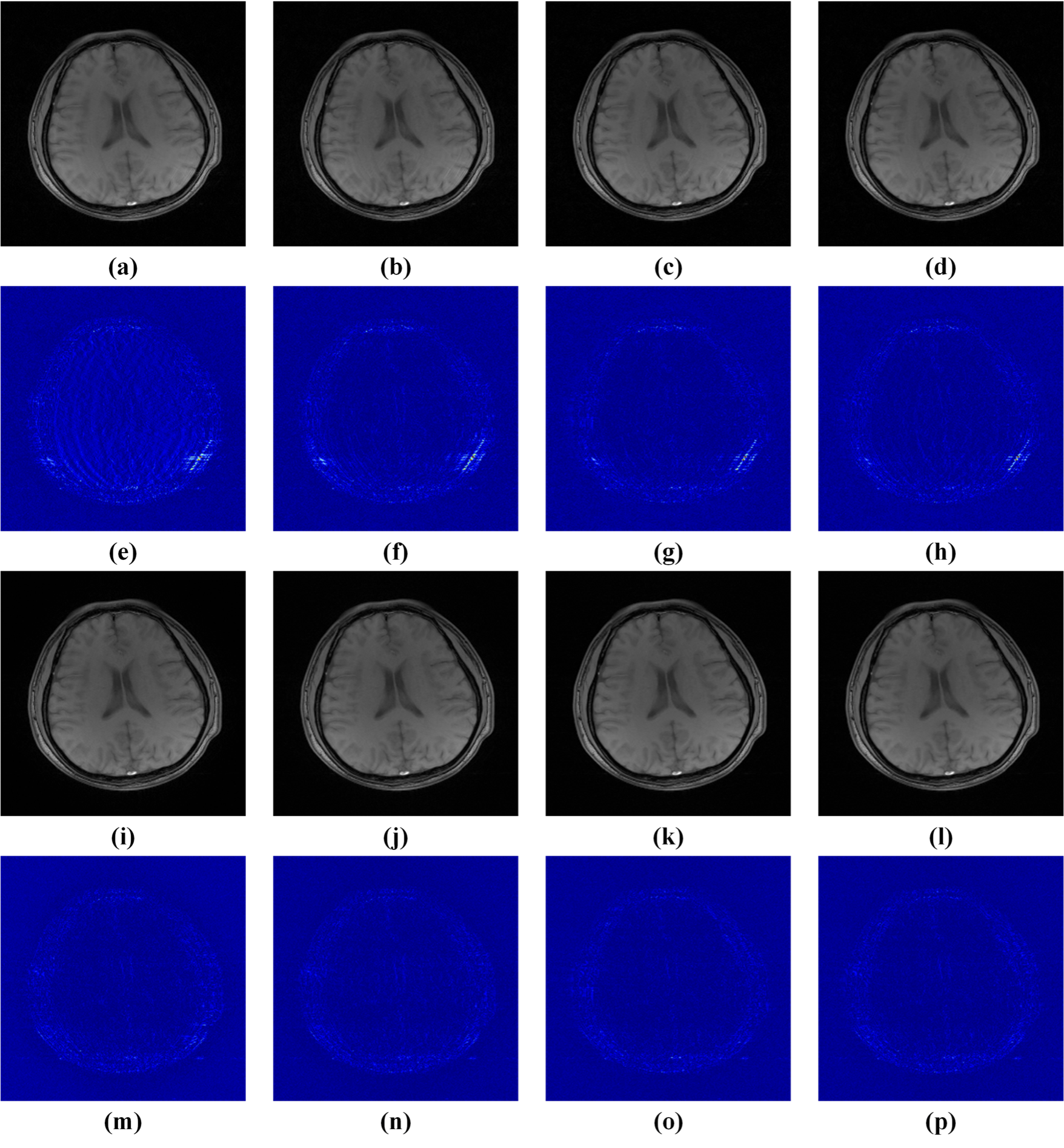}
\caption{Reconstruction results of four tight frames under a sampling rate of $0.34$. (a-d) (or (i-l)) are images of reconstruction results of pFISTA-SENSE with $\gamma=1$ (or pFISTA-SPIRiT with $\gamma = 1/c$) using contourlet, shearlet, PBDW, and PBDWS, respectively; (e-h) and (m-p) are the reconstruction error distribution (3x) corresponding to reconstructed image above them. Note: 8-coil image in Fig. 3 (a) in the main text and sampling pattern in Fig. 3 (d) in the main text are adopted in all experiments.}
\label{fig_Recon34}
\end{figure}

Since the sampling rate of $0.25$ is relatively low for the Cartesian sampling pattern, the reconstructed images are somewhat blurred and appear artifacts in Fig. \ref{fig_psi_images}. Therefore, we conducted experiments at a higher sampling rate of $0.34$, and the results are presented in Fig. \ref{fig_Convergence34} and \ref{fig_Recon34}. As can be seen, when the sampled data are relatively adequate, RLNEs drop to a much lower level (Fig. \ref{fig_Convergence34}). Notably, the results indicate the effectiveness of the proven convergence criteria. The shearlet, PBDW, and PBDWS permit nice reconstructed images with excellent artifacts suppression. In contrast, contourlet produces images appearing slightly blur (Fig. \ref{fig_Recon34}).


%----------------------------------------------------------------------
%----------------------------- References -----------------------------
%----------------------------------------------------------------------
\ifCLASSOPTIONcaptionsoff
  \newpage
\fi

% \newpage
% \bibliographystyle{IEEEtran}
% \bibliography{IEEEabrv,Mylib}